\documentclass[a4paper,11pt]{article}
\pdfoutput=1 
\usepackage{verbatim}

\usepackage{jheppub}
\usepackage{amsmath,amsfonts,amssymb,amsthm,graphicx,color}
\usepackage{slashbox}
\usepackage[T1]{fontenc} 
\usepackage{enumerate}
\usepackage{tikz, tikz-3dplot, pgfplots}
\usepackage[utf8]{inputenc}
\usepackage{slashed}
\usepackage{subcaption}

\newcommand{\ba}[1]{\begin{align} #1 \end{align} }

\newcommand{\bs}[1]{\begin{split} #1 \end{split} }
\def\tr{\text{Tr}}
\def\CN{{\cal N}}
\def\CS{{\cal S}}
\def\CI{{\cal I}}
\newcommand{\mc}[1]{\mathcal{ #1} }

\usepackage{romannum}
\usepackage{genyoungtabtikz}

\title{\boldmath 	Structure of Anomalies of 4d SCFTs from M5-branes, and Anomaly Inflow}

\author[a]{Ibrahima Bah}
\author[b]{and Emily Nardoni} 
\affiliation[a]{Department of Physics and Astronomy, Johns Hopkins University, 3400 North Charles Street, Baltimore, MD 21218, USA}
\affiliation[b]{Department of Physics, University of California, San
Diego, La Jolla, CA 92093, USA}

\emailAdd{iboubah@jhu.edu, enardoni@ucsd.edu}

\abstract{We study the 't Hooft anomalies of four-dimensional superconformal field theories that arise from M5-branes wrapped on a punctured Riemann surface.  In general there are two independent contributions to the anomalies.  There is a bulk term obtained by integrating the anomaly polynomial of the world-volume theory on the M5-branes over the Riemann surface; this contribution knows about the punctures only through its dependence on the Euler characteristic of the surface. The second set of contributions comes from local data at the punctures; these terms are independent from the bulk data of the surface. Using anomaly inflow in M-theory, we describe the general structure of the anomalies for cases when the four-dimensional theories preserve $\CN=2$ supersymmetry.  We additionally discuss the anomalies corresponding to $(p,q)$ punctures in $\CN=1$ theories.  }

\preprint{}

\begin{document} 

\maketitle
\flushbottom

\newpage

\section{Introduction}

't Hooft anomalies provide a robust measure of the degrees of freedom in quantum field theories. 
In general, the anomalies for a given theory in even spacetime dimensions $d$ can be encoded in a $(d+2)$-form polynomial known as the anomaly polynomial, which depends on  the various curvature forms associated to the dynamical or background gauge and gravity fields. If the gauge or gravity field is dynamical, its anomalies should vanish or else the theory is inconsistent. Otherwise, the anomaly doesn't lead to an inconsistency, but often has interesting physical consequences. We refer to anomalies in background gauge or gravity fields as 't Hooft anomalies. For a review, see \cite{AlvarezGaume:1984dr,Harvey:2005it}.    

In the last ten years, there has been a proliferation of new classes of four-dimensional Superconformal Field Theories (SCFT's), dubbed class $\mathcal{S}$, that are inherently strongly coupled and admit no known Lagrangian description.  These theories can emerge from the low energy limit of six-dimensional $\CN=(2,0)$ SCFT's wrapped on a punctured Riemann surface, which in certain cases describe the low energy dynamics of M5-branes.  These constructions have been revolutionary in that they provide a partial classification scheme for four-dimensional $\mathcal{N}=2$ SCFT's, and bring to bear new geometric tools for studying them \cite{Gaiotto:2009we,Gaiotto:2009gz,Gaiotto:2009hg}.  Soon after their introduction, it was demonstrated that these constructions can be generalized to study the space of four-dimensional $\mathcal{N}=1$ SCFT's \cite{Maruyoshi:2009uk,Benini:2009mz,Bah:2011je,Bah:2011vv,Bah:2012dg,Beem:2012yn}.

The basic set-up of class $\mathcal{S}$ theories is reviewed in Section \ref{sec:classS}.  An important ingredient in the construction is a partial topological twist \cite{Witten:1988ze,Bershadsky:1995qy}, which is needed to preserve some supercharges in the compactification of the six-dimensional theory.  Depending on the choice of twist, various amount of supersymmetry can be preserved in four dimensions.  

Anomalies are particularly important observables for the theories of class $\CS$, as they provide a measure of various degrees of freedom in these inherently strongly coupled field theories. The anomalies for $\mathcal{N}=2$ class $\mathcal{S}$ theories can be obtained, in some cases, by using S-duality \cite{Gaiotto:2009we,Gaiotto:2009gz,Chacaltana:2010ks} and anomaly matching on the moduli space \cite{Chacaltana:2012zy,Tachikawa:2015bga}. In the special case of $\mathcal{N}=1$ theories we can use Seiberg duality as well as anomaly matching on the moduli space to obtain them \cite{Gadde:2013fma,Bah:2013aha,Agarwal:2014rua}.  

In the cases where they are known, the anomaly polynomials of class $\CS$ theories have two contributions which are independent and must be stated separately. The first is the contribution from the bulk Riemann surface, which we denote $I_6(\Sigma_{g,n})$. This depends only on the genus $g$ and number of punctures $n$ through the Euler characteristic $2g-2+n$, and on the anomaly polynomial of the UV six-dimensional theory. The second set of contributions come from new degrees of freedom localized at the punctures; these are related to the consistent boundary conditions for the six-dimensional theory at these locations.  A contribution of a puncture to the anomalies of the four-dimensional theory is denoted as $I_6(P)$. The total six-form anomaly polynomial $I_6^{\CS}$ for the class $\CS$ theory takes the form
	\ba{
I_6^{\CS}=I_6(\Sigma_{g,n}) + \sum_{i=1}^n I_6(P_i).\label{eq:cs1}
	}
	
Since the theories of class $\mathcal{S}$ are defined by the compactification of a six-dimensional theory, there should exist a prescription for directly computing their anomaly polynomials from the geometric construction.  Indeed, in the case of theories obtained by compactifying on a smooth Riemann surface without punctures, integrating the anomaly polynomial $I_8$ of the six-dimensional theory over the surface can yield the polynomial of the four-dimensional theory \cite{Alday:2009qq,Bah:2012dg}\footnote{This procedure fails when there are accidental symmetries.  This problem is most commonly encountered when compactifying on a Riemann surface with vanishing Euler characteristic---see for example \cite{Bah:2017gph}.}, i.e. 
	\ba{
		I_6^{\mathcal{S}} = \int_{\Sigma_{g}}I_8.\label{eq:integrate1}
	}
	This prescription requires shifting the curvature of the background R-symmetry gauge field with the curvature form of the Riemann surface, implementing the topological twist.  The integration of the eight-form polynomial over the surface picks out the terms that are linear in the surface's curvature form, and therefore proportional to its volume form.  
	
In the presence of punctures this prescription fails; we cannot obtain the full anomaly polynomial of class $\mathcal{S}$ theories by simply shifting the background curvature and integrating.  There can be additional terms in the anomaly polynomial of the six-dimensional theory, and the integration over the Riemann surface cannot account for the additional data localized at the punctures.  
 
 Our primary goal in this paper is to discuss the anomaly polynomial in Class $\CS$ theories, and make the separation \eqref{eq:cs1} between the bulk and puncture pieces explicit. We outline a prescription for computing the anomaly polynomial using inflow, and argue that this general form \eqref{eq:cs1} follows from anomaly inflow in M-theory on the M5-branes wrapping the punctured Riemann surface. 
 
 Our strategy is motivated by the holographic duals of class $\mathcal{S}$ theories from punctured Riemann surfaces \cite{Gaiotto:2009gz,Bah:2015fwa} (see \cite{Bah:2013wda} for probe analysis).  In the gravity duals, the topological twists are manifested by non-trivial $S^1$-bundles over the Riemann surface.  The connections on these bundles are related to the shifts of the background R-symmetry in the twist, and their curvatures $F$ are proportional to the volume form of the Riemann surface.  In the presence of a puncture, $F$ picks up monopole sources that encode the new degrees of freedom associated to the puncture.  These monopoles are end points of additional M5-branes localized at the puncture and extended along a direction normal to the surface.  This gives a strong hint that in computing $I_6^\mathcal{S}$, we need to enrich the shifting prescription of the background gauge field of the R-symmetry to account for these sources.  Moreover, in integrating the eight-form anomaly polynomial, there is an additional interval that is normal to the branes along which the monopole sources sit.  We indeed recover these features in the anomaly inflow analysis.

 In our analysis, we allow for localized monopole sources in the curvature form $F$ over the Riemann surface. These sources will cause the four-form M-theory flux  to pick up new terms. Accounting for this, we argue that the inflow procedure yields additional contributions to the anomalies of the world-volume theory that depend on the puncture data.


The structure of the rest of this paper is as follows. In Section \ref{sec:classS} we set the stage for the rest of the paper. We review the construction of four-dimensional field theories preserving $\CN=2$ or $\CN=1$ supersymmetry from M5-branes wrapped on Riemann surfaces, and describe general features of their anomalies. Readers familiar with the class $\CS$ construction and anomalies can skip to Section \ref{sec:n1}. 

In Section \ref{sec:n1} we take a field-theory detour. Focusing on the case where the four-dimensional field theory preserves $\CN=1$ supersymmetry, we derive the anomalies corresponding to a large class of locally $\CN=2$ preserving punctures in geometries in which the bulk preserves $\CN=1$. We additionally discuss an illuminating way of parameterizing the anomaly coefficients we obtain in terms of an $\CN=1$ generalization of an effective number of vector multiplets and hypermultiplets. This section and the inflow discussion that follows may be read independently of one another. 

In the remainder of the paper, we discuss the class $\CS$ anomalies by anomaly inflow in 11d supergravity in the presence of M5-branes, focusing on the case where the four-dimensional theory preserves $\CN=2$. We begin with a review of inflow for flat M5-branes in Section \ref{sec:inflow1} (originally discussed in \cite{Freed:1998tg,Harvey:1998bx,Witten:1996hc}). In Section \ref{sec:inflow2} we discuss the anomaly eight-form of the M5-branes in the curved background, and argue for the general structure observed from the field theory analysis as described in Section \ref{sec:classS}.

In \cite{Bah:2018bn}, we discuss more details of the inflow analysis.

\section{Structure of Class $\mathcal{S}$ Anomalies} \label{sec:classS}

This section serves as an extended introduction to the four-dimensional theories obtained by compactifying the six-dimensional (2,0) theories on a Riemann surface, setting notation and focusing attention on the main points of interest in the rest of the paper. The experienced reader can skip to Section \ref{sec:n1}.

\subsection{The class $\CS$ construction}
\label{sec:construction}

A large class of four-dimensional quantum  field theories can be studied by compactifying six-dimensional $\CN=(2,0)$ superconformal field theories over a punctured Riemann surface with a partial topological twist. The SCFTs that result from this procedure are known as theories of class $\CS$. Generically, these theories are strongly coupled and do not have a known Lagrangian description, and yet many of their properties can be inferred by utilizing their origin in six dimensions and the compactification scheme.

Depending on the choice of twist, the low energy four-dimensional theory can preserve various amounts of supersymmetry.  $\CN=2$ theories of class $\CS$ were first constructed and classified in \cite{Gaiotto:2009we, Gaiotto:2009hg} (building on earlier work by \cite{Witten:1997sc}).
A large class of $\CN=1$ SCFTs and their dualities were studied via mass deformations of $\CN=2$ theories in \cite{Maruyoshi:2009uk,Benini:2009mz,Bah:2011je}. 
Later, it was demonstrated that $\CN=1$ SCFTs could be directly constructed from compactifications of six-dimensional theories on a Riemann surface with a partial topological twist \cite{Bah:2012dg,Bah:2011vv}. A strong piece of evidence for the existence of these superconformal theories is the explicit construction of their large-$N$ gravity duals. The gravity duals for the $\CN=2$ theories corresponding to M5-branes wrapped on Riemann surfaces without punctures were constructed in \cite{Gaiotto:2009gz}, which were holographically dual to the Maldacena-Nu{\'n}ez supergravity solutions \cite{Maldacena:2000mw}. The duals for the $\CN=1$ theories were constructed in \cite{Bah:2012dg,Bah:2011vv} (without punctures) and in \cite{Bah:2015fwa} (with punctures).

Generically, putting a QFT on a curved background breaks supersymmetry. A partial topological twist allows us to preserve some supersymmetry in the IR. In the twist, one turns on a background gauge field valued in the six-dimensional $SO(5)_R$ symmetry, and tunes it to cancel the background curvature on the Riemann surface. We identify an abelian subgroup of the six-dimensional R-symmetry as
	\ba{
	U(1)_+\times U(1)_-\subset SU(2)_+ \times SU(2)_-\subset SO(5)_R,\label{eq:symm}
	}
where the $U(1)_\pm$ are Cartans of the $SO(5)_R$. Then, embed the holonomy group of the Riemann surface $U(1)_h$ in the six-dimensional R-symmetry group by identifying the $U(1)_h$ generator $R_h$ as a linear combination of the $U(1)_\pm$ generators $J_\pm$, 
	\ba{
	R_h=\frac{p_1}{p_1+p_2}J_+ + \frac{p_2}{p_1+p_2}J_-. \label{eq:embed}
	}
This fixes the parameters $(p_1,p_2)$ in terms of the Euler characteristic $\chi$ of the surface as
	\ba{
	p_1+p_2+\chi(\Sigma_{g,n})=0,\qquad \text{with}\qquad -\chi(\Sigma_{g,n}) = 2(g-1)+n.\label{eq:conditiona}
	}
This procedure in general preserves four supercharges in four dimensions, and breaks the bosonic symmetries of the six-dimensional theory as
	\ba{
	SO(1,5)\times SO(5)_R\to SO(1,3)\times U(1)_+\times U(1)_-.\label{eq:breaking}
	}
When one of $(p_1,p_2)$ is zero, eight supercharges will be preserved, and one of the $U(1)_\pm$ will be enhanced to $SU(2)_\pm$ to furnish the $\CN=2$ R-symmetry of the four-dimensional theory.

The six-dimensional $(2,0)$ theories are labeled by a choice of gauge algebra $\mathfrak{g}$, which follows an ADE classification. The $\mathfrak{su}(N)=A_{N-1}$ and $\mathfrak{so}(2N)=D_{N}$ cases have a description in terms of M5-branes. In the present work we will focus on the $A_{N-1}$ theories, in which case the six-dimensional theory arises as the effective world-volume theory of $N$ coincident M5-branes. Then, the amount of supersymmetry that is preserved in the IR depends on the way the M5-branes are embedded in a Calabi-Yau threefold $CY_3$. From this perspective, the Riemann surface is described by a holomorphic curve $\mc{C}_{g,n}$ in $CY_3$. Generally, the Calabi-Yau threefold is a $U(2)$ bundle over $\mc{C}_{g,n}$, whose determinant line bundle is fixed to the canonical bundle of the surface. We choose to twist the Cartan of the $SU(2)$ bundle, such that the $U(2)$ bundle decomposes into a sum of two line bundles $\mc{L}_1$ and $\mc{L}_2$ with integer degrees $(p_1,p_2)$ as\footnote{Often these parameters are called $(p,q)$ in the literature, but here we reserve the $(p,q)$ labels to specify the R-symmetry of locally $\CN=2$ preserving punctures.}
	\ba{ \bs{
	\begin{tikzpicture}
	\node at (0,0) {$\mathbb{C}^2$};
	\draw[->] (0.4,0) -- (1,0);
	\node at (1.8,0) {$\mc{L}_1\oplus \mc{L}_2$};
	\draw[->] (1.8,-.3) -- (1.8,-1);
	\node at (1.8,-1.3) {$\mc{C}_{g,n}$};  \label{eq:cygeometry} 
	\end{tikzpicture} }}
	
In this language, the topological twist involves embedding the holonomy group $U(1)_h$ of the Riemann surface in the $SO(5)$ structure group of the normal bundle to the M5-branes. The $U(1)_\pm$ global symmetries in (\ref{eq:symm}) correspond to phase rotations of the two line bundles. Requiring that the first Chern class of the Calabi-Yau threefold vanish is equivalent to the condition \eqref{eq:conditiona}. If one of the two line bundles is trivial, the threefold decomposes as $CY_3=CY_2\times \mathbb{C}$, and the background preserves eight supercharges.\footnote{In the special case that the Riemann surface is a 2-torus ($g=1$), $\CN=4$ supersymmetry can be preserved by fixing the normal bundle to the M5-brane world-volume to be trivial.}

The four-dimensional theories of class $\CS$ also have a description in terms of a generalized quiver gauge theory. To take the $A_{N-1}$ case, one geometrically  decomposes the curve $\mc{C}_{g,n}$ into 3-punctured spheres connected by tubes via pair-of-pants decompositions. S-duality relates different degeneration limits of the curve. The low energy effective description of $N$ coincident M5-branes wrapping a 3-punctured sphere is known as the $T_N$ theory, which is a strongly coupled $\CN=2$ SCFT with an $SU(N)^3$ global symmetry \cite{Gaiotto:2009we}. Gauging subgroups of these global symmetries via $\CN=1$ or $\CN=2$ vector multiplets (the tubes) corresponds geometrically to ``gluing'' punctures to form Riemann surfaces with general Euler characteristic. The classification of these four-dimensional building blocks, or ``tinkertoys'', has been carried out in \cite{Chacaltana:2010ks, Chacaltana:2011ze, Chacaltana:2014jba, Chacaltana:2017boe,Chcaltana:2018zag}. Similar field-theoretic constructions have recently been obtained for theories whose geometries have negative line bundle degrees $p_1$ and $p_2$ \cite{Agarwal:2015vla,Fazzi:2016eec,Nardoni:2016ffl}. Such a field-theoretic approach can be useful in providing a different perspective on the properties of the theories of class $\CS$.

Much of the richness of class $\CS$ comes from the punctures on the Riemann surface. From the perspective of the parent six-dimensional $(2,0)$ theory, punctures are 1/2-BPS codimension-2 defects,  specified by an embedding $\rho:\mathfrak{su}(2)\to \mathfrak{g}$. Such embeddings are labeled by nilpotent orbits of the Lie algebra $\mathfrak{g}$ \cite{Chacaltana:2012zy}. Regular defects correspond to singular boundary conditions to Hitchin's equation on the Riemann surface, which have been classified for $A_{N-1}$ in \cite{Gaiotto:2009gz}, for $D_N$ in \cite{Tachikawa:2009rb}, and have been discussed for other types of regular defects including twisted lines (possible when the ADE group admits an outer-automorphism) in \cite{Tachikawa:2010vg, Chacaltana:2012zy,Chacaltana:2013oka,Chacaltana:2015bna}. The generalization to $\CN=1$ Hitchin's equations was first discussed in \cite{Xie:2013gma}. In the present work, we will only discuss regular (also called ``tame'') defects, and omit discussion of irregular (``wild'') punctures corresponding to higher order poles. 

When the (2,0) theories describe the effective world-volume theory of M5-branes, punctures correspond to points where the M5-branes branch out to infinity \cite{Bah:2013qya}. 
The punctures correspond to boundaries of the Riemann surface, and boundary conditions are needed for the M5-branes; this leads to global symmetries. 

At large $N$, one can look at AdS$_5$ dual solutions of M-theory corresponding to the near horizon limit of $N$ M5-branes wrapping a Riemann surface \cite{Gaiotto:2009gz,Bah:2015fwa}. In these solutions, the new degrees of freedom are associated to additional M5-branes that are localized at the punctures (see  \cite{Bah:2013qya}). These branes are extended along a direction normal to the Riemann surface, and end at monopole sources of a $U(1)$ connection of an $S^1$ bundle over the surface.  This connection is associated to the topological twist in the field theory construction. A single M5-brane corresponds to a simple puncture, which can be analyzed in the probe approximation \cite{Gaiotto:2009gz,Bah:2013wda}. In the full backreacted solution, the connection forms in the Ricci flat background pick up monopole sources. 

To summarize, we require the following data in order to specify a theory of class $\CS$: a choice of $\mathfrak{g}=$ADE; the Euler characteristic $\chi$ of the Riemann surface; a choice of twist, i.e. the $(p_1,p_2)$ that satisfy (\ref{eq:conditiona}); and local data associated with the punctures.
From this perspective, the class $\CS$ construction allows us to organize a large space of four-dimensional SCFTs in a geometric way.

\subsection{Anomalies of the $(2,0)$ theories}
\label{sec:anomalies20}

The six-dimensional $\CN=(2,0)$ theories are labeled by an ADE Lie algebra: $A_{N-1}=\mathfrak{su}(N),D_{N}=\mathfrak{so}(2N)$, or $\mathfrak{e}_{6,7,8}$. The six-dimensional (2,0) superconformal algebra is $\mathfrak{osp}(4|8)$. Its bosonic subgroup is $SO(2,6)\times  SO(5)_R$, corresponding respectively to the conformal group and R-symmetry group.
These theories arise from decoupling limits of string theory constructions \cite{Witten:1995zh,Strominger:1995ac,Witten:1995em}. 

The focus of the present work is the anomalies of the four-dimensional class $\CS$ theories, which can be understood by tracking the anomaly polynomial of their parent six-dimensional theories in the compactification on the surface. The six-dimensional (2,0) theories cannot be written down in terms of the usual path integral of local fields, which makes understanding their properties a challenge. However, as anomalies are inherently topological quantities, they are accessible even for these mysterious theories. 

The interacting $A_{N-1}$ theory is the effective world-volume theory of $N$ coincident M5-branes, and the $D_N$-type theories are realized on the world-volume of $N$ coincident M5-branes at an $\mathbb{R}^5/\mathbb{Z}_2$ orbifold fixed point. In these cases, the derivation of six-dimensional $(2,0)$ anomalies can be understood in terms of inflow for M5-branes in 11d supergravity. As the M5-brane world-volume is six-dimensional, the anomalies will involve eight-dimensional characteristic classes, packaged in an eight-form anomaly polynomial which encodes anomalous diffeomorphisms of the world-volume of the M5-branes and their normal bundle. The idea of the inflow analysis is that in the presence of the M5-branes, the total anomaly from zero modes on the world-volume and inflow from the bulk should vanish in order for the theory to be consistent. Using these methods (which we review in Section \ref{sec:inflow1}), the anomaly eight-form for a single M5-brane is derived as \cite{Witten:1996hc, Freed:1998tg}
	\ba{
	I_8[1]= \frac{1}{48} \left[ p_2(NW)-p_2(TW) + \frac{1}{4} \left( p_1(TW) - p_1(NW) \right)^2\right]. \label{eq:m51}
	}
$NW$ and $TW$ are the normal bundle and tangent bundle to the M5-brane world-volume $W$, respectively, and $p_k$ are the Pontryagin classes, reviewed in Appendix \ref{sec:anomalypolynomials}.  (\ref{eq:m51}) is also the anomaly polynomial for a single, free $(2,0)$ tensor multiplet. The tensor multiplet is the only (2,0) superconformal multiplet that describes free fields, containing a self-dual three-form, as well as Weyl fermions in the spinor representation of $SO(5)_R$, and real scalars in the fundamental of $SO(5)_R$.

	\begin{table}[t!]
	\centering
	\begin{tabular}{|c|c|c|c|}
	\hline 
	$G$ & $r_G$ & $d_G$ & $h_G$ \\ \hline 
	$A_{N-1}$ & $N-1$ & $N^2-1$ & $N$ \\
	$D_{N}$ & $N$ & $N(2N-1)$ & $2N-2$ \\
	$E_6$ & $6$ & $78$ & $12$ \\
	$E_7$ & $7$ & $133$ & $18$ \\
	$E_8$ & $8$ & $248$ & $30$ \\ \hline
	\end{tabular}
	\caption{Rank, dimension, and Coxeter numbers for the simply-laced Lie groups. Note the useful group theory identity $d_G=r_G(h_G+1)$.\label{tab:numbers}}
	\end{table}

For a general six-dimensional (2,0) theory of type $\mathfrak{g}=\ $ADE, the eight-form anomaly polynomial takes the form
	\ba{
	I_8[\mathfrak{g}] = r_G I_8[1] + \frac{d_Gh_G}{24} p_2(NW).\label{eq:i8g}
	}
The values of $r_G,d_G,h_G$ for the ADE groups are listed for reference in Table \ref{tab:numbers}. Here, the normal bundle $NW$ can be thought of as an $SO(5)$ bundle coupled to the six-dimensional R-symmetry. 

This result was obtained for $A_{N-1}$ in \cite{Harvey:1998bx} via inflow with multiple M5-branes, and conjectured for all $\mathfrak{g}=\ $ADE in  \cite{Intriligator:2000eq} using purely field-theoretic reasoning. It was verified for $D_N$ in \cite{Yi:2001bz} with an inflow analysis, and verified for all $\mathfrak{g}=$ADE in \cite{Ohmori:2014kda} via anomaly matching on the tensor branch.  An exact calculation of the $a$-anomaly for (2,0) theories via a similar field-theoretic derivation was given in  \cite{Cordova:2015vwa}. The famous $N^3$ scaling at large $N$ was first noticed in the context of black hole calculations of the thermal free energy \cite{Klebanov:1996un}, and was computed for the central charges via AdS/CFT \cite{Henningson:1998gx}.

\subsection{Structure of class $\CS$ anomalies}
\label{sec:gens}

As we emphasized in the introduction, the anomalies of class $\CS$ have two contributions which are independent and must be stated separately:
	\ba{
	I_6^{\CS}=I_6(\Sigma_{g,n}) + \sum_{i=1}^n I_6(P_i).\label{eq:cs}
	}
Here, we'll give a more complete discussion of this point.

The bulk piece $I_6(\Sigma_{g,n})$ is always obtained by integrating the eight-form anomaly polynomial (\ref{eq:i8g}) over the Riemann surface with a given Euler characteristic $\chi=-2g+2-n$, and with the appropriate topological twist (\ref{eq:conditiona}), as in \eqref{eq:integrate1} \cite{Bah:2012dg,Bah:2011vv,Alday:2009qq}.
This piece will be proportional to $\chi$, since the terms in $I_8[\mathfrak{g}]$ that survive the integral are linear in the curvature two-form on the Riemann surface.

	\begin{figure}[t!]
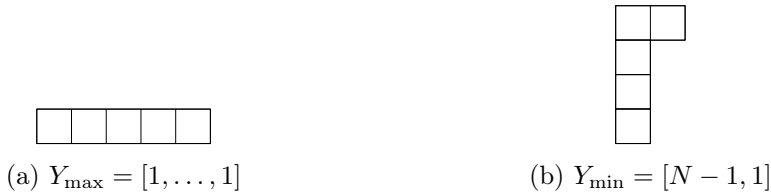

	\centering
	\subcaptionbox{$Y_{\text{max}}=[1,\dots,1]$}%
	  [.45\linewidth]{\yng(5)}
	\subcaptionbox{$Y_{\text{min}}=[N-1,1]$}
	  [.45\linewidth]{\yng(2,1,1,1)}
 	 \caption{A maximal puncture (left) corresponds to a Young diagram with $N$ columns of height 1. A minimal puncture (right) corresponds to a Young diagram with one column of height $N-1$ and one column of height 1.\label{fig:puncs}}
	\end{figure}
	
The second class of terms are due to the punctures. Deriving these contributions from a six-dimensional perspective is more subtle. These pieces depend on local data which add degrees of freedom to the theory, leading to global symmetries. In this note we'll be interested in the anomalies of a class of punctures dubbed regular punctures, which we review below. For more details on the anomalies of regular punctures, see Appendix \ref{sec:punctures}.

A regular puncture is labeled by an embedding $\rho: \mathfrak{su}(2)\to \mathfrak{g}$. For $\mathfrak{g}=A_{N-1}$, the choice of embedding is 1-to-1 with a partition of $N$, and is therefore labeled by a Young diagram $Y$. For a Young diagram with $n_i$ columns of height $h_i$, the field theory will have an unbroken flavor symmetry $G$ associated to the puncture. $G$ corresponds to the commutant of the embedding $\rho$, given as
	\ba{
	G=S\left[\prod_i U(n_i)\right]. \label{eq:flavorp}
	}
The case of the maximal flavor symmetry $G=SU(N)$ is known as a maximal (or full) puncture, and the case of the minimal flavor symmetry $G=U(1)$ is known as a minimal (or simple) puncture. These limiting cases correspond to the Young diagrams given in Figure \ref{fig:puncs}.

The form of $I_6(P_i)$ can be derived from string dualities utilizing the generalized quiver descriptions of the four-dimensional theories \cite{Chacaltana:2012zy}.  One can derive the contributions to the anomaly polynomial from non-maximal punctures by Higgsing the associated flavor symmetry and keeping track of the multiplets which decouple \cite{Tachikawa:2015bga}.

The additive structure of the anomalies (\ref{eq:cs}) is motivated by the TQFT structure of the class $\CS$ theories. Both the $\CN=1$ and $\CN=2$ class $\CS$ SCFTs admit a formulation in the language of a 2d topological quantum field theory \cite{Gadde:2009kb,Gadde:2011uv,Gadde:2011ik,Gaiotto:2012xa,Beem:2012yn}. The superconformal index is then computed as an $n$-point correlation function of the TQFT living on the $n$-punctured Riemann surface, with punctures corresponding to operator insertions. Thus, the theories are organized topologically by specifying bulk information and local puncture information. 

One should note, however, that even though the anomaly polynomial has a simple additive structure, quantities of interest such as the central charges are still nontrivial and don't follow immediately from topological arguments. For instance, in the $\CN=1$ case there is an additional $U(1)$ flavor symmetry that mixes with the $U(1)_R$ symmetry. Given an R-symmetry, the anomaly polynomial encodes all the mixed anomalies with the global symmetries. However, $a$-maximization is required to specify the exact superconformal R-symmetry.

\subsection*{Anomalies for $\CN=2$ SCFTs}

Here, we review the anomalies for the four-dimensional $\CN=2$ SCFTs which we will match onto in an inflow computation in Section \ref{sec:inflow2}.

The anomaly polynomial of a four-dimensional $\CN=2$ superconformal theory with a flavor symmetry $G$ and $SU(2)_R\times U(1)_R$ symmetry has the form\footnote{More generally, the term $k_G c_1(F_{1}) c_2(F_G)$ should be written in terms of the instanton number $n(F_G)$, normalized such that for $SU(N)$ $n(F_{SU(N)})=c_2(F_{SU(N)})$---e.g. see \cite{Tachikawa:2015bga}.}
	\ba{
	I_6=& (n_v-n_h) \left( \frac{c_1 (F_{1})^3}{3} - \frac{c_1(F_{1})p_1(T^4)}{12} \right) - n_v c_1(F_{1}) c_2(F_{2}) + k_G c_1(F_{1}) c_2(F_G). \label{eq:i6n2}
	}
This expression follows from the definition of the anomaly polynomial for four-dimensional Weyl fermions, as reviewed in Appendix \ref{sec:anomalypolynomials}, and the $\CN=2$ superconformal algebra \cite{Kuzenko:1999pi}. 
In \eqref{eq:i6n2}, $F_{1}$ ($F_{2}$) is the field strength for the background gauge field of the $U(1)_R$ ($SU(2)_R$) bundle, and $F_G$ is the field strength of the flavor symmetry bundle. The flavor central charge $k_G$ is defined in Appendix \ref{sec:scfts}. More generally, additional flavor symmetries would contribute additional terms in \eqref{eq:i6n2}. 

The parameters $n_v$ and $n_h$ are related to the central charges of the SCFT as $a=\frac{1}{24}(5n_v+n_h)$, and $c=\frac{1}{12}(2n_v+n_h)$. If the theory is free, then $n_v$ and $n_h$ denote the number of vector multiplets and hypermultiplets respectively; otherwise, we regard $n_v$ and $n_h$ as an effective number of vector and hypermultiplets. Even for interacting field theories, this notation serves as a useful bookkeeping device.

The R-symmetry of the $\CN=2$ theories described in Section \ref{sec:construction} is identified as $SU(2)_+\times U(1)_-$ when $p_1=0$, and as $U(1)_+\times SU(2)_-$ when $p_2$ is zero. Denoting the generators of the $U(1)_\pm$ as $J_\pm$ and the $SU(2)_R\times U(1)_R$ generators by $I^a$ and $R_{\CN=2}$ respectively (see Appendix \ref{sec:scfts} for conventions), this corresponds to the identification
	\ba{\bs{
	p_1&=0:\qquad J_+=2I^3,\qquad J_-=R_{\CN=2}\\
	p_2&=0:\qquad J_+=R_{\CN=2},\qquad J_-=2I^3.
	}\label{eq:gens}}

As summarized in (\ref{eq:cs}), the theories of class $\CS$ have two contributions to their anomalies: contributions from the bulk, and local contributions from the punctures. For the $\CN=2$ theories, as suggested by \cite{Gaiotto:2009gz} it is convenient to write these in terms of an effective number of vector and hypermultiplets $(n_v,n_h)$ as
	\ba{
	n_v=n_v(\Sigma_{g,n}) + \sum_{i=1}^n n_v(P_i),\qquad n_h=n_h(\Sigma_{g,n}) + \sum_{i=1}^n n_h(P_i).\label{eq:nvnh}
	}
These terms were computed explicitly in \cite{Gaiotto:2009gz,Chacaltana:2010ks}, with the help of a result in \cite{Nanopoulos:2010ga}. The bulk terms are given by\footnote{Note that in much of the literature, the term proportional to the $n$ in $\chi=-(2g-2+n)$ is instead grouped with the puncture contribution to the anomalies. The grouping we use here emphasizes the fact that the whole term proportional to $\chi$ comes from global considerations. E.g. regardless of the types of punctures, this term only depends on their total number.}
	\ba{
	n_v(\Sigma_{g,n})= -\frac{\chi}{2} \left(r_G + \frac{4}{3}d_Gh_G\right),\qquad n_h(\Sigma_{g,n}) = -\frac{\chi}{2} \left(\frac{4}{3}d_Gh_G\right).\label{eq:nvback}
	}
The puncture contributions $n_{v,h}(P_i)$ for the $A_{N-1}$ case are reviewed in Appendix \ref{sec:punctures}. As detailed there, these terms depend on the details of the Young diagrams corresponding to the punctures. 

Together, the bulk contribution (\ref{eq:nvback}), and the puncture contributions (\ref{eq:nvy}) and (\ref{eq:nhy}) determine the full anomaly polynomial of the four-dimensional $\CN=2$ class $\CS$ SCFTs. Plugging into \eqref{eq:i6n2}, this gives
	\ba{
	I_6(\Sigma_{g,n})&=-\frac{\chi}{2} \left[ \left(\frac{\left(c_1^+\right)^3}{3} - \frac{c_1^+p_1(T^4)}{12}\right) r_G - c_1^+c_2^- \left(r_G + \frac{4}{3} d_G h_G\right)\right],\label{eq:ibulka}\\ \bs{
	I_6(P_i)&= \left(n_v(P_i)-n_h(P_i)\right) \left(\frac{(c_1^+)^3}{3} - \frac{c_1^+p_1(T^4)}{12}\right)  - n_v(P_i) c_1^+c_2^- + k_{G_i} c_1^+{c_2({F_{G_i}})}. }\label{eq:ipunc}
	}
Here, we've chosen $p_2=0$ as our $\CN=2$ limit, with $c_1^+\equiv c_1(U(1)_+)$ and $c_2^-\equiv c_2(SU(2)_-)$. The terms proportional to $(c_1^+)^3$ and $c_1^+c_2^-$ are 't Hooft anomalies for the background R-symmetry. The $c_1^+p_1(T^4)$ pieces encode the mixed gauge-gravity anomalies. The last piece in $I_6(P_i)$ couples the global symmetry $G_i$ preserved by the puncture with the $U(1)_R$ symmetry, and will have a separate term for each factor in the puncture flavor symmetry (\ref{eq:flavorp}). The anomaly coefficients $\tr R_{\CN=2}=\tr R_{\CN=2}^3$ and $\tr R_{\CN=2}I_3^2$ are readily determined from (\ref{eq:ipunc}) using (\ref{eq:n2coef}) and (\ref{eq:n2kcoef}).

\subsubsection*{General structure of $\CN=1$ class $\CS$ anomalies}

The $\CN=1$ theories of class $\CS$ preserve a $U(1)_+\times U(1)_-$ global symmetry which derives from the $\CN=(2,0)$ $SO(5)_R$ symmetry as in (\ref{eq:symm}). 
A combination of the $U(1)_\pm$ generators $J_\pm$ corresponds to a flavor symmetry $\mc{F}$, and we can pick an R-symmetry $R_0$, given as  
	\ba{
	R_0=\frac{1}{2}(J^++J^-),\qquad \mc{F}=\frac{1}{2}(J^+-J^-).\label{eq:r0f}
	}
The exact superconformal R-symmetry $R_{\CN=1}$ is 
	\ba{
	R_{\CN=1}(\epsilon)=R_0+\epsilon \mc{F}=\frac{1}{2} (J_+ + J_-) - \frac{1}{2} \epsilon (J_+ - J_-),\label{eq:sr}
	}
where $\epsilon$ is determined by $a$-maximization \cite{Intriligator:2003jj}. When $p_1=0$, $\epsilon$ is fixed to be $\frac{1}{3}$, and the generators $J_\pm$ are identified as in \eqref{eq:gens}. In this case, we identify an $\CN=1$ subalgebra in $\CN=2$ as
	\ba{
	R_{\CN=1}=\frac{1}{3}R_{\CN=2} + \frac{4}{3}I^3 = \frac{1}{3} J_+ + \frac{2}{3}J_-,
	}
and $U(1)_-$ is enhanced to $SU(2)_-$. Similarly, when $p_2=0$, $\epsilon=-\frac{1}{3}$ and $U(1)_+$ is enhanced to $SU(2)_+$.

The 't Hooft anomalies for the four-dimensional class $\CS$ theories are encoded in a six-form anomaly polynomial. It follows from \cite{Anselmi:1997am} and the definition of the anomaly polynomial discussed in Appendix \ref{sec:anomalypolynomials} that the anomaly polynomial for a four-dimensional theory with a $U(1)_+\times U(1)_-$ global symmetry takes the form
	\begin{equation}
	I_6^{\CS} = \frac{1}{6} \tr \left[ J_+ c_1^+ + J_- c_1^-\right]^3 - \frac{1}{24} \tr \left[ J_+ c_1^+ + J_- c_1^-\right]p_1(T^4).   \label{eq:i6n1b}
	\end{equation}
Here, $c_1^\pm \equiv c_1(U(1)_\pm)$ are the first Chern classes of the $U(1)_\pm$ bundles. 
There could be additional flavor symmetries, which will mix with the R-symmetry and give additional terms in \eqref{eq:i6n1b}.
	
As discussed in Section \ref{sec:gens}, the anomaly polynomials for the $\CN=1$ class $\CS$ theories will decompose into background contributions from the bulk which can be computed directly by integrating $I_8[\mathfrak{g}]$ for the six-dimensional theory over the Riemann surface, and local contributions from the punctures. The bulk contribution to $I_6^{\CS}$ is  
	\ba{\bs{
	I_6(\Sigma_{g,n})=-\frac{\chi(1+z)}{2} \left\{ \left(\frac{\left(c_1^+\right)^3}{6}-\frac{c_1^+p_1(T^4)}{24}\right)r_G - \frac{c_1^+\left(c_1^-\right)^2}{2}\left( r_G + \frac{4}{3}d_Gh_G\right) \right\}\\
	-\frac{\chi(1-z)}{2} \left\{ \left(\frac{\left( c_1^-\right)^3}{6}-\frac{c_1^-p_1(T^4)}{24}\right)r_G - \frac{c_1^-\left(c_1^+\right)^2}{2}\left( r_G + \frac{4}{3}d_Gh_G\right) \right\}.
	}}
We've written the answer in terms of the twist parameter $z$, defined
	\ba{
	z=\frac{p_1-p_2}{p_1+p_2},\qquad p_1+p_2=2g-2+n=-\chi(\Sigma_{g,n}).\label{eq:twist}
	}	
This result for the bulk anomalies follows from the analysis in \cite{Bah:2012dg}. In the next section, we'll give a discussion of the contributions of punctures to the anomalies of the $\CN=1$ class $\CS$ theories.

\section{Anomalies of $(p,q)$ Punctures in Class $\mathcal{S}$} \label{sec:n1}

In this section, we study the anomalies of a large class of allowed punctures in the  $\CN=1$ class $\CS$ SCFTs which carry a $(p,q)$ label. We present new results for the anomalies of $(p,q)$ punctures. This section can be read independently of sections \ref{sec:inflow1} and \ref{sec:inflow2}.

\subsection{Anomalies of $(p,q)$ punctures}
\label{sec:pqs}

When the bulk preserves $\CN=1$, there are punctures that can locally preserve $\CN=2$ supersymmetry. In this case the local degrees of freedom preserve $\CN=2$ supersymmetry, and therefore there is a local $\CN=2$ R-symmetry action. This action is identified with the background $J_\pm$ symmetries in a nontrivial way, with different choices labeled by $(p,q)$. For a given background with fixed $J_\pm$, there is an infinite family of inequivalent $(p,q)$-labeled punctures. The existence of these $(p,q)$ punctures has been demonstrated in the gravity duals \cite{Bah:2015fwa}, with the $(p,q)$ restricted to co-prime integers, but as of yet they have not been understood in general from a field theory perspective.

We identify the generators of the $SU(2)_R\times U(1)_R$ symmetry locally near a $(p,q)$ puncture as 
	\begin{equation}
	R_{\mathcal{N}=2} = \frac{p}{p-q} J_+ - \frac{q}{p-q} J_-, \qquad 2 I_3 = \frac{q}{q-p} J_+ - \frac{p}{q-p} J_-.\label{eq:pqpunc}
	\end{equation}  
Once $R_{\mathcal{N}=2}$ is fixed as a general linear combination of $J_\pm$, we can fix $I_3$ by identifying the flavor symmetry $(R_{\mathcal{N}=2} -2I_3)$ with the combination $(J_+ -J_-)$. Then, we can refine the statement of Section \ref{sec:construction} of what local data is required to specify a $\CN=1$ theory of class $\CS$. When the Riemann surface has punctures that locally preserve $\CN=2$, one must specify:  
		\begin{itemize}
		\item A choice of embedding $\rho: \mathfrak{su}(2)\to \mathfrak{g} =$ADE, determining the flavor symmetry at the puncture, and
		\item A choice of $(p,q)$, determining the R-symmetry locally at the puncture as (\ref{eq:pqpunc}).
		\end{itemize}

	The $(p,q)$ punctures are a generalization of the notion of ``colored'' punctures that appear in field theory, e.g. in \cite{Bah:2013aha,Gadde:2013fma,Agarwal:2014rua,Giacomelli:2014rna}. Punctures in theories in which the bulk spacetime preserves $\CN=1$ supersymmetry have an additional $\mathbb{Z}_2$-valued label $\sigma=\pm1$ called the ``color'', which corresponds to the fact that we can choose one of the two normal directions to the M5-branes at the location of the puncture. In the gravity dual, the puncture corresponds to D4 branes ending on D6 branes, and the choice of $\sigma$ corresponds to the choice of a plane transverse to the D4's along which the D6's are extended. In the more general framework of $(p,q)$ punctures, these choices correspond to
		\ba{\bs{
		\sigma=+1\qquad&\leftrightarrow\qquad (p,q)=(p,0)\\
		\sigma=-1\qquad&\leftrightarrow\qquad (p,q)=(0,q).
		}\label{eq:color}}
	For $\sigma=+1$ the geometry locally preserves a $U(1)_+\times SU(2)_-$ bundle, while for $\sigma=-1$ a $U(1)_-\times SU(2)_+$ bundle is preserved. The overall normalization in (\ref{eq:pqpunc}) was fixed by matching onto these two limiting cases.

For general $(p,q)$ punctures, the anomaly coefficients can be computed with the local twist (\ref{eq:pqpunc}). We'll express the answer for the anomaly coefficients in terms of a local twist parameter $\hat{z}$, defined analogously to (\ref{eq:twist}) as
	\ba{
	\hat{z}=\frac{p-q}{p+q}.\label{eq:twist2}
	}
When $q=0$, $\hat{z}=1$, and when $p=0$, $\hat{z}=-1$, such that $\hat{z}$ reduces to the $\sigma=\pm 1$ label in these limits. The result is that a puncture corresponding to a flavor symmetry $G$ with a $(p,q)$ twist yields the following contribution to the anomaly polynomial of the four-dimensional theory:
	\ba{\bs{
	I_6(P_i,\hat{z}) &= (1+\hat{z}) \left[ {a_+^{(1)}} {\left(c_1^+\right)^3} - {a_+^{(2)}}{c_1^+p_1(T^4)} - {a_+^{(3)}} {c_1^+ \left(c_1^-\right)^2}+ \frac{k_G}{3} c_1^+c_2(F_G) \right]\\
	&+(1-\hat{z})\left[ a_-^{(1)} {\left(c_1^-\right)^3} - a_-^{(2)} {c_1^-p_1(T^4)} -a_-^{(3)} {c_1^-\left(c_1^+\right)^2} + \frac{k_G}{3} c_1^-c_2(F_G) \right].
	}\label{eq:i6pq}}
The coefficients $a_+^{(i)}$ are given by
	\ba{\bs{
	a_+^{(1)} &= -\frac{1}{24} \left( n_h(P_i)(1+\hat{z})^2 + 2 n_v(P_i) (1-4\hat{z}+\hat{z}^2) \right) \\
	a_+^{(2)} &=-\frac{1}{24} (n_v(P_i)-n_h(P_i)),\qquad a_+^{(3)} =\frac{1}{8} \left( n_h(P_i)(1-\hat{z})^2 + 2n_v(P_i)(1+\hat{z}^2) \right),
	}}
and $a_-^{(i)}(\hat{z})=a_+^{(i)}(-\hat{z})$. The effective number of vector multiplets $n_v(P_i)$ and hypermultiplets $n_h(P_i)$ in the $A_{N-1}$ case are given in (\ref{eq:nvy}) and (\ref{eq:nhy}). The flavor central charge terms are as given in \eqref{eq:i6pq} for our current definition of $(p,q)$ punctures, but one can imagine a case where the $c_2(F_G)$ also splits. The rules for this splitting are not clear, and will not be further discussed here.

$I_6(P_i,\hat{z})$ reduces to the answer already known for punctures with $\hat{z}=\pm 1$. For example, the contribution of a maximal puncture with $\hat{z}=-1$ in the $A_{N -1}$ case reduces to
	\ba{
	I_6(P_{\text{max}},1)&= -\frac{(N^2-1)}{2} \left( \frac{\left(c_1^+\right)^3}{3} - \frac{c_1^+p_1(T^4)}{12} - c_1^+\left(c_1^-\right)^2\right) + 2 Nc_1^+c_2(F_{SU(N)}),
	}
	which matches $n_v(P_{\text{max}})=-d_G/2$ and $n_h(P_{\text{max}})=0$.

\subsection{Effective $n_v,n_h$ for $\CN=1$ theories}
\label{sec:effective}

It was conjectured in \cite{Agarwal:2014rua} that even when the bulk doesn't preserve $\CN=2$ supersymmetry, an $\CN=1$ analogue of $n_v$ and $n_h$ can be defined. In this section, we check this proposal for $\CN=1$ class $\CS$ theories with general $(p,q)$ punctures. To do so, we'll use the convenient basis of $(R_0,\mc{F})$ defined in (\ref{eq:r0f}) for the four-dimensional global symmetries. When the $\CN=1$ theory is derived from an $\CN=2$ theory, $R_0$ and $\mc{F}$ are related to the $\CN=2$ R-symmetry generators as $R_0=R_{\CN=2}/2+I_3,\ \mc{F}=-R_{\CN=2}/2+I_3$. It will be further useful to express results in terms of the twist parameters $z$ and $\hat{z}$, as defined in (\ref{eq:twist}) and (\ref{eq:twist2}).

The proposal of \cite{Agarwal:2014rua} is that the anomaly coefficients for the $\CN=1$ class $\CS$ theories can be written in the form
\ba{\bs{
\tr R_0=n_v-n_h,\qquad \tr R_0^3=n_v-\frac{n_h}{4},\\
\tr \mc{F}=-(\hat{n}_v-\hat{n}_h),\qquad \tr \mc{F}^3=-\hat{n}_v+\frac{\hat{n}_h}{4},\\
\tr R_0 \mc{F}^2 = -\frac{n_h}{4},\qquad \tr R_0^2 \mc{F}=\frac{\hat{n}_h}{4}.
}\label{eq:conjecture}}
These parameters are defined such that when it is possible to identify the $\CN=1$ subalgebra as part of an $\CN=2$ algebra, $n_v$ and $n_h$ are precisely the effective number of free vector and hypermultiplets of the $\CN=2$ theory. The $\hat{n}_v$ and $\hat{n}_h$ parameters are defined analogously for the anomalies involving an odd power of the flavor symmetry $\mc{F}$, and should be loosely interpreted as the $\CN=1$ version of an effective number vector and hypermultiplets.

We find that (\ref{eq:conjecture}) is true only for $\hat{z}=\pm 1$ punctures, and requires some modification for more general $(p,q)$ punctures.  Writing the anomalies in terms of  $(n_v,\hat{n}_v,n_h,\hat{n}_h)$, we find that the $\CN=1$ class $\CS$ anomaly coefficients take the form
\ba{\bs{
\tr R_0 &= n_v-n_h,\qquad\qquad\qquad\qquad\quad\  \tr F = -\left( \hat{n}_v-\hat{n}_h\right),\\
\tr R_0^3 &= - \frac{n_h}{4} + n_v + \frac{3}{2} \sum_i\delta_-(P_i),\ \quad\tr F^3 =\frac{\hat{n}_h}{4} - \hat{n}_v + \frac{1}{2}\sum_i\hat{z_i}\delta_+(P_i), \\
\tr R_0F^2&=-\frac{n_h}{4} ,\qquad\qquad\qquad\qquad\quad\   \tr FR_0^2=\frac{\hat{n}_h}{4}.
}\label{eq:actual}}
Relative to the $\CN=2$ class $\mc{S}$ theories, the $n_v$ and $n_h$ parameters have additional local terms. They break into separate bulk and local contributions as
	\ba{
	n_{v,h}&=n_{v,h}(\Sigma_{g,n}) + \sum_{i=1}^n \left[n_{v,h}(P_i)+\delta_-(P_i)\right],\label{eq:nvh}\\
	\hat{n}_{v,h}&=-z n_{v,h}(\Sigma_{g,n}) -\sum_{i=1}^{n} \hat{z}_i \left[n_{v,h}(P_i)+\delta_+(P_i)\right]. \label{eq:nhatvh}
	}
The bulk pieces $n_{v,h}(\Sigma_{g,n})$ are the same as in the $\CN=2$ case, which we repeat here for clarity:
	\ba{
	n_v(\Sigma_{g,n})= -\frac{\chi}{2} \left(r_G + \frac{4}{3}d_Gh_G\right),\qquad n_h(\Sigma_{g,n}) = -\frac{\chi}{2} \left(\frac{4}{3}d_Gh_G\right).
	}
The local pieces $n_{v,h}(P_i)$ are reviewed in Appendix \ref{sec:punctures} for the $A_{N-1}$ case, and depend on the data of the Young diagram associated to the flavor symmetry at the puncture. In effect, the parameters $\hat{n}_{v,h}$ special to the $\CN=1$ theories are a twisted version of the $\CN=2$ parameters.
	
The deviation from the conjecture (\ref{eq:conjecture}) lies in the $\delta_\pm(P_i)$ terms, which written in terms of the $(p,q)$ parameters are given as
	\ba{
	\delta_\pm(P_i) \equiv \frac{2pq}{(p+q)^2}\left(n_h(P_i)\pm 2 n_v(P_i) \right).
	}
When either $p$ or $q$ is zero, (\ref{eq:actual}) matches onto (\ref{eq:conjecture}), and we recover the known answer for the effective $(n_v,n_h,\hat{n}_v,\hat{n}_h)$. Otherwise, these represent new contributions to the anomalies. For nonzero $p$ and $q$, they contribute extra effective vector multiplets and hypermultiplets to the theory that depend on local puncture data, due to their appearance in (\ref{eq:nvh}) and (\ref{eq:nhatvh}). Additionally, they contribute new terms to the cubic anomalies, such that (\ref{eq:actual}) deviates from (\ref{eq:conjecture}). This result can be stated as the fact that we require more than four parameters to label the anomalies of theories with $(p,q)$ punctures.

\subsubsection*{Discussion}

For the moment, let's get some intuition as to the meaning of the $(n_v,n_h,\hat{n}_v,\hat{n}_h)$ parameters for the cases where $\delta_\pm(P_i)=0$. Consider two class $\CS$ theories that each have an $SU(N)$ flavor symmetry. By gauging a diagonal subgroup of the $SU(N)\times SU(N)$ symmetries with either an $\CN=1$ or $\CN=2$ vector multiplet, we glue the two punctures associated with the flavor symmetries. Then, we can isolate the contribution of the $\CN=1$ or $\CN=2$ vector multiplet to the anomalies as \cite{Agarwal:2014rua}
	\ba{\bs{
\CN=1\ \text{vector}&:\quad	n_v=N^2-1,\quad n_h=0,\quad \hat{n}_v=0,\quad\quad \hat{n}_h=0\\
\CN=2\ \text{vector}&:\quad	n_v=N^2-1,\quad n_h=0,\quad \hat{n}_v=N^2-1,\quad \hat{n}_h=0.}
} 
These precisely correspond to subtracting the contributions of two maximal punctures of different colors (for $\CN=1$ gluing) or of the same color (for $\CN=2$ gluing).

Another simple example is to consider the $A_{N-1}$ (2,0) theory compactified on a sphere with two maximal punctures and one minimal puncture. This is the $T_N$ theory with one puncture partially closed, and corresponds to the theory of $N^2$ free hypermultiplets $H^i=(Q^i,\tilde{Q}^i),\ i=1,\dots,N$ in the bifundamental representation of the $SU(N)\times SU(N)$ flavor symmetry. For instance, with $N=2$ the theory is Lagrangian, and one can explicitly check that the matter content is four $\CN=2$ hypermultiplets, or eight $\CN=1$ chiral multiplets. The contribution of these hypermultiplets to the anomaly is 
	\ba{
(N,\bar{N}) \text{ hypermultiplets}&:\quad n_v=0,\quad   n_h=N^2,\quad \hat{n}_v=0,\quad \hat{n}_h=N^2,
	}
as expected. In both of these cases, the $\hat{n}_{v,h}$ parameters have a natural interpretation in terms of splitting the $\CN=2$ multiplets into $\CN=1$ components. 

As a final example, consider the case when the Riemann surface has $n^{(+)}$ maximal punctures with $\hat{z}=1$, and $n^{(-)}$ maximal punctures with $\hat{z}=-1$. Denote the total number of punctures as $n_{tot}=n^{(+)}+n^{(-)}$, and let $n_{dif}=n^{(-)}-n^{(+)}$. Then, $(n_v,\hat{n}_v,n_h,\hat{n}_h)$ reduce to the known results (see e.g. \cite{Nardoni:2016ffl})
	\ba{\bs{
	n_v=-\frac{\chi}{2} (r_G + \frac{4}{3} d_G h_G)- \frac{d_G}{2}n_{tot},\qquad n_h=-\frac{2\chi}{3} d_Gh_G,\\
	\hat{n}_v=\frac{z\chi}{2}  (r_G + \frac{4}{3} d_G h_G)- \frac{d_G}{2} n_{dif},\qquad \hat{n}_h=\frac{2z\chi}{3}d_Gh_G.
	}}
	
In all of these cases, there is an interpretation of the $(n_v,n_h,\hat{n}_v,\hat{n}_h)$ parameters in the generalized quiver description of the $\CN=1$ theory. The generalized quiver description is also useful in understanding colored punctures with non-maximal flavor symmetry by Higgsing an operator in the adjoint of the flavor symmetry group with a nilpotent vev, as discussed in the context of $\CN=2$ theories in \cite{Tachikawa:2013kta}, and in the context of $\CN=1$ class $\CS$ theories in \cite{Agarwal:2014rua}. For general $(p,q)$ punctures with $\delta_\pm\neq0$, however, we do not currently have a field theory interpretation in terms of a generalized quiver. The fact that the anomalies for the $(p,q)$ punctures take the form (\ref{eq:actual}) implies that there is no straightforward field-theoretic interpretation of gluing $(p,q)$ punctures. It would be further interesting to understand the operation of closing maximal punctures via nilpotent Higgsing from the perspective of the anomaly polynomial for the $\CN=1$ class $\CS$ theories, as was discussed for the $\CN=2$ theories in \cite{Tachikawa:2015bga}. The additional $U(1)$ symmetry in the $\CN=1$ case that mixes with the R-symmetry naively complicates the problem. We leave these interesting questions to upcoming work.

\section{Inflow for Flat M5-branes: A Review} \label{sec:inflow1}

The anomalies of the (2,0) theories of type $A_{N-1}$ and $D_N$ can be obtained by inflow in 11d supergravity in the presence of M5-branes. The eight-form anomaly polynomial \eqref{eq:i8g} encodes anomalous diffeomorphisms of the six-dimensional world-volume of the M5-branes and their normal bundle.  Here, we will restrict our attention to the $A_{N-1}$ case, for which the four-dimensional class $\CS$ theories have a description as the low energy limit of $N$ coincident M5-branes wrapped on a punctured Riemann surface. Our goal will be to describe the inflow procedure for this class of theories. Before we get there, we will take some time to review the standard inflow mechanism for M5-branes. In Section \ref{sec:inflow2} we will extend this analysis to the main problem of interest.

\subsection{Anomaly inflow}\label{sec:inflow}

A QFT that admits chiral fields coupled to gauge or gravity fields may have anomalies. In even spacetime dimensions $d$, consistent anomalies are encoded in a $(d+2)$-form $I_{d+2}$ known as the anomaly polynomial. $I_{d+2}$ is a polynomial in the dynamical or background gauge and gravity fields\footnote{The anomaly polynomial can also involve differential forms on the space of couplings of the theory, as was recently pointed out in the context of class $\CS$ theories in \cite{Tachikawa:2017aux}.}, and is related to the anomalous variation of the quantum effective action as
	\ba{
	\delta {S}_{\text{eff}} = 2\pi \int_{M_d} I_{d}^{(1)}. \label{eq:variation1}
	}
Here, $I_{d}^{(1)}$ is a $d$-form obtained from $I_{d+2}$ via the descent procedure \cite{AlvarezGaume:1983ig,AlvarezGaume:1984dr,Bardeen:1984pm},
	\ba{
	I_{d+2}=dI_{d+1}^{(0)},\qquad \delta I_{d+1}^{(0)} = dI_{d}^{(1)}.
	}
$\delta$ indicates the gauge variation, and the superscripts indicate the order of the quantity in the gauge variation parameter.

In string theory, gauge theories can be obtained by considering the decoupling limit of extended objects---such as branes---in a gravitational background. Gauge transformations and/or diffeomorphisms restricted on the branes induce global symmetries. If effective degrees of freedom of the world-volume theory on the branes are chiral (possible when the world-volume is even-dimensional), then the induced global symmetries can be anomalous.  Since diffeomorphisms in the full gravitational theory must be preserved, the action of the gravitational theory in the presence of the brane sources must be anomalous in order to cancel the anomalies of the world-volume theory. 

In inflow, the anomaly is canceled by a term in the bulk effective action whose variation is localized on the brane \cite{Faddeev:1985iz,Callan:1984sa}. Such a coupling implies a source in the equations of motion, modifying the Bianchi identity for the $(D-p-1)$-form field strength $d H_{D-p-1} = \delta_{D-p}$ (for $D$-dimensional spacetime). The anomalous variation of the effective action can be written in terms of the descent of a $(p+3)$-form anomaly polynomial $I_{p+3}$ as in \eqref{eq:variation1}, where the integral will be over the $(p+1)$-dimensional world-volume. This procedure was first explained in \cite{Callan:1984sa}, while a detailed study of the role played by consistent versus covariant anomalies appeared in \cite{Naculich:1987ci}. An extension to Green-Schwarz anomaly cancellation appeared in \cite{Blum:1993yd}.  Such anomalous terms in the presence of Dp-branes were understood in \cite{Green:1996dd,Cheung:1997az,Minasian:1997mm}. For a review of D-brane and I-brane (intersecting D-brane) inflow, including an extended discussion on regularizing the delta function sources in this context, see \cite{Kim:2012wc}.

In the context of M-theory, the (5+1)-dimensional M5-brane carries a chiral tensor multiplet, which has a one-loop anomaly; this is canceled by inflow from the bulk. 
The origin of the anomaly in M-theory  comes from topological terms in the supergravity action, which have an anomalous variation in the presence of the M5-branes. Because the M5-brane acts as a magnetic source for the $C_3$ potential of M-theory,  inflow can be understood as a result of the modified Bianchi identity (schematically) $dG_4=\delta_5$. For a nice review of anomaly cancellation in M-theory, see \cite{Bilal:2003es}.

\subsection{M5-brane inflow}
\label{sec:inflowreview}

Now, we review the inflow analysis for flat M5-branes in 11d supergravity. Anomaly inflow for a single flat M5-brane was first discussed in \cite{Duff:1995wd}, and the computation was done in \cite{Witten:1996hc} and \cite{Freed:1998tg}.  Inflow for $N$ flat M5-branes was computed in \cite{Harvey:1998bx}. We will use the details and notation reviewed in this section as a jumping off point in Section \ref{sec:inflow2}.

The eleven-dimensional supergravity action is given by
	\ba{
	S = \frac{1}{2\kappa_{11}^2} \int \sqrt{-g} \left( R - \frac{1}{2} |G_4|^2 \right)- \frac{1}{12 \kappa_{11}^2} \int C_3\wedge G_4 \wedge G_4 -\mu_{M_2} \int C_3\wedge I_8^{\text{inf}}.\label{eq:action}
	}
$C_3$ is the three-form gauge field, $G_4=dC_3$ the four-form field strength, and $\mu_{M_2}$ the M2-brane tension. The integrals are over eleven-dimensional spacetime, $M_{11}$. 

The couplings satisfy
	\begin{equation}
	\frac{1}{2\kappa_{11}^2} = \frac{2\pi}{(2\pi \ell_p)^9}, \qquad \mu_{M_2} = \frac{2\pi}{(2\pi \ell_p)^3}, \qquad \mu_{M_5} = \frac{2\pi}{(2\pi \ell_p)^6},
\end{equation} 
and we fix $2\pi \ell_p =1$ such that $\mu_{M_2}=\mu_{M_5}=2\pi$. The eight-form $I_8^{\text{inf}}$ is a polynomial function of the spacetime curvature $R$ on $M_{11}$, 
	\begin{equation}
	I_8^{\text{inf}} = - \frac{1}{48} \left( p_2(R) - \frac{1}{4} \left(p_1(R)\right)^2 \right),\label{eq:iinf}
	\end{equation}
with conventions for the Pontryagin classes given in Appendix \ref{sec:anomalypolynomials}.

Diffeomorphisms in the bulk are anomalous in the presence of M5-branes. For an M5-brane with six-dimensional world-volume $W_6$, the tangent bundle to $M_{11}$ splits as
	\ba{
	TM_{11}|_{W_6} = TW_6 \oplus NW_6, \label{eq:tmsplit}
	}
with $TW_6$ an $SO(1,5)$ bundle, and $NW_6$ an $SO(5)$ bundle.  Diffeomorphisms of $M_{11}$ that map $W_6\to W_6$ induce $SO(1,5)$ diffeomorphisms of the world-volume (gravitational anomalies) and $SO(5)$ gauge transformations of the normal bundle (gauge anomalies).  

The M5-branes magnetically source the four-form flux $G_4$, modifying the Bianchi identity for $G_4$ as\footnote{The source appears with units given as $
dG_4=2\kappa_{11}^2 \mu_{M5} N \delta_{5}$; in units where $2\pi \ell_p=1$, $2\kappa_{11}^2 \mu_{M5}=1$. }
	\begin{equation}
	dG_4= N \delta_{5}, \qquad  \delta_5 = \delta(y^1)\dots \delta(y^5)dy^1\wedge\dots \wedge dy^5. \label{eq:source} 
	\end{equation}
Here, the $y^a$ coordinates parameterize the transverse space to the M5-branes, which sit at $y^a=0$. Terms in the bulk action \eqref{eq:action} are singular due to \eqref{eq:source}, leading to inflow towards the world-volume that should be canceled by anomalies carried by degrees of freedom on the M5-branes. 
A proper treatment requires that we smooth out the delta functions at the positions of the M5-branes  \cite{Freed:1998tg}. We will need to replace the delta functions with bump functions, and impose regularity and gauge invariance of the field strength. This will imply a particular form of the gauge transformation of $C_3$. For this discussion we restrict to the case of a single brane.

To implement the smoothing of the source, parameterize the transverse directions to the M5-brane by an $S^4$ whose volume form is 
	\ba{\bs{
	dV_5 = \left(\frac{1}{4!} \epsilon_{abcde} d\hat{y}^a\wedge d\hat{y}^b\wedge d \hat{y}^c \wedge d\hat{y}^d \hat{y^e}\right)\wedge r^4 dr\equiv  d\Omega_4\wedge r^4\ dr.
	}}
	The $\hat{y}^a$ are isotropic coordinates on the $S^4$ fibers of the sphere bundle over the M5-brane world-volume. We smear the charge over the radial direction with a smooth function $\rho(r)$ that satisfies $\rho(0)=-1$ and $\rho(r\to\infty)=0$, such that the Bianchi identity \eqref{eq:source} is
	\ba{
	dG_4= d\rho(r) \wedge e_4. \label{eq:thom}
	}
The four-form $e_4$ is a closed, global angular form that is gauge invariant under $SO(5)$ transformations of the normal bundle and restricts to $d\Omega_4$ when the $SO(5)$ connection is taken to be trivial. Denoting the $SO(5)$ gauge field as $F^{ab}$, $e_4$ takes the form
	\ba{\bs{
	e_4 &= \frac{1}{V_4} \left( D\Omega_4 - \frac{2}{4!} \epsilon_{abcde}F^{ab}\wedge D\hat{y}^c \wedge D\hat{y}^d\hat{y}^e + \frac{1}{4!}\epsilon_{abcde}F^{ab}\wedge F^{cd}\hat{y}^e \right).
	}\label{eq:e4so5}}
Here $D\Omega_4$ refers to $d\Omega_4$ with ordinary derivatives replaced with covariant derivatives, and $V_4$ refers to the area of the $S^4$, $V_4=8\pi^2/3$. This is normalized such that integrating $e_4$ over the $S^4$ gives unity\footnote{Our conventions in this section follow \cite{Freed:1998tg}, except that their normalization is such that $\int_{S_4}e_4^{\text{them}}=2$.}. Note that $de_4=0$ ensures that $d^2G_4=0$.

\eqref{eq:thom} requires that the relationship between $G_4$ and $C_3$ is modified to
	\ba{
	G_4=dC_3-d\rho \wedge e_3^{(0)}.
	}
Here, $e_3^{(0)}$ is a two-form related to $e_4$ by the standard descent procedure,
	\ba{
	e_4=de_3^{(0)},\quad \delta e_3^{(0)}=de_2^{(1)}.
	}
Requiring gauge invariance of $G_4$ then implies that $C_3$ has an anomalous variation,
	\ba{
	\delta C_3=-d\rho\wedge e_2^{(1)}.\label{eq:gaugec3}
	}

In order to compute the variation of the action in the presence of the M5-branes, $G_4$ and $C_3$ need to replaced with quantities that are smooth and non-singular in the neighborhood of the branes.  It follows from \eqref{eq:gaugec3} that the correct replacement is
\begin{equation}
C_3 \; \to \; C_3 - \rho e_3^{(0)}, \qquad G_4 \; \to \; d \left(C_3 - \rho e_3^{(0)} \right).  \label{eq:shift}
\end{equation}

Now we have the pieces to compute the variation of the bulk action and the anomaly due to the brane source. There are two terms in the bulk action that can lead to an anomaly: the linear coupling $C_3\wedge I_8^{\text{inf}}$ (Green-Schwarz), and the $C_3\wedge G_4\wedge G_4$ (Chern-Simons) terms. From the decomposition of the tangent bundle \eqref{eq:tmsplit}, it follows that $I_8^{\text{inf}}$ can be written as
	\ba{
	I_8^{\text{inf}} = -\frac{1}{48} \left[ p_2(TW_6) +p_2(NW_6) -\frac{1}{4} \left(p_1(TW_6)-p_1(NW_6)\right)^2\right].
	}
The variation of $I_8^{\text{inf}}$ is given by the descent formalism as 
\begin{equation}
	I_8^{\inf} = d I_7^{\text{inf}(0)}, \qquad \delta I_7^{\text{inf}(0)} = d I_6^{\text{inf}(1)}.
	\end{equation} 	

We will need to regulate the integrals by removing a neighborhood of radius $\epsilon$ around the M5-brane. Denote by $D_\epsilon(W_6)$ the total space of the disc bundle with base $W_6$ and with fibers the discs of radius $\epsilon$. First one computes the variation outside the disc with the shifted non-singular forms \eqref{eq:shift}. Then, take the size of the disc to zero. The total space of the $S^4$ sphere bundle over $W_6$ which forms the boundary of $M_{11}/D_{\epsilon}(W_6)$ will be denoted as $S_\epsilon(W_6)$. 
Then, the variation of the linear term leads to
\begin{align}
	\frac{\delta S_{L}}{2\pi } &=-\lim_{\epsilon\to 0} \int_{M_{11}/D_\epsilon(W_6)}  d\rho \wedge e_4 \wedge I_6^{\text{inf}(1)} = \int_{W_6}I_6^{\text{inf}(1)}. \label{eq:varlin}
\end{align}
The other source of anomalies in the bulk action is the Chern-Simons term, improved to take into account \eqref{eq:shift} as
	\ba{
	\frac{S_{CS}'}{2\pi}&=-{\frac{1}{6}}\lim_{\epsilon\to 0} \int_{M_{11}/D_\epsilon(W_6)} \left(C_3 -\rho e_3^{(0)} \right) \wedge \left(G_4 -\rho e_4 \right) \wedge \left(G_4 -\rho e_4 \right).  \label{eq:modifiedc}
	}
Its variation leads to
	\ba{
	\frac{\delta S_{CS}'}{2\pi} = \frac{1}{{6}} \int_{S_\epsilon(W_6)} e_2^{(1)}\wedge e_4 \wedge e_4   =  \frac{1}{24} \int_{W_6} \left[p_2(NW_6)\right]^{(1)}\equiv \int_{W_6} I_6^{CS(1)}, \label{eq:var}
	}
where the second equality is due to a result of Bott-Catteneo \cite{bott1999integral}, and $\left[p_2(NW_6)\right]^{(1)}$ refers to the six-form related to $p_2(NW_6)$ by descent. In both \eqref{eq:varlin} and \eqref{eq:var} we have integrated by parts.

Combining this contribution with the contribution from the $C_3\wedge I_8^{\text{inf}}$ term, the anomaly eight-form for a single M5-brane is then
	\ba{
	I_8[1]=I_8^{\text{inf}} + I_8^{CS} = \frac{1}{48} \left[ p_2(NW_6)-p_2(TW_6) + \frac{1}{4} \left( p_1(TW_6) - p_1(NW_6) \right)^2\right]. \label{eq:m52}
	}
This is precisely the result we quoted in \eqref{eq:m51}. For $N$ M5-branes, the Green-Schwarz term is linear in $C_3$ and thus also linear in $N$, and the Chern-Simons term is cubic in $C_3$ and thus also cubic in $N$ (we take $\rho(r=0)=-N$). Anomaly cancellation then requires that $I_8[N]$ for $N$ M5-branes is given by 
	\ba{
	I_8[N] &= I_8^{CS}[N] + I_8^{\text{inf}}[N]= (N^3-N) \frac{p_2(NW_6)}{24} + N I_8[1]. \label{eq:I8flat}
	}
To obtain the anomaly polynomial for the six-dimensional $A_{N-1}$ theories, we must also subtract off an overall $U(1)$ corresponding to the center of mass motion of the branes, which amounts to subtracting $I_8[1]$ from \eqref{eq:I8flat}.

\section{Class $\CS$ Anomalies from Inflow} \label{sec:inflow2}

We turn to the problem of anomaly inflow for cases where the M5-branes wrap a holomorphic curve $\mathcal{C}_{g,n}$. 
As we emphasized in the previous section, anomaly inflow in a gravitational theory can be understood as accounting for localized sources in variations of the action. The logic we will employ is as follows.  We can account for boundary conditions for the M5-branes by considering additional sources that model the branching off of the M5-branes at the punctures. This is consistent with the M-theory description of punctures as transverse M5-branes that intersect the Riemann surface at a point. Such sources will yield additional terms in the the anomalous variation of the action above.

\subsection{Inflow for curved M5-branes}

As we reviewed in Section \ref{sec:inflowreview}, in the presence of an M5-brane the tangent bundle to the full eleven-dimensional spacetime splits into the tangent bundle and normal bundle to the world-volume, as \eqref{eq:tmsplit}. When the M5-branes wrap a holomorphic curve $\mathcal{C}_{g,n}$, the tangent bundle over the branes further splits as
\begin{equation}
TW_6 = TM^{1,3} \oplus T\mathcal{C}_{g,n}.
\end{equation} Since the Riemann surface is embedded in a $CY_3$ that is a sum of two line bundles, the $SO(5)$ normal bundle over the branes reduces to a sum of two $SO(2)$ bundles,
\begin{equation}
NW_6 = SO(2)_+ \oplus SO(2)_-.
\end{equation} 
We can consider the $U(1)_+\times U(1)_-$ cover of the normal bundle on the flat four-dimensional space, which correspond to the global symmetries in the field theory in \eqref{eq:breaking}. We will reduce the curvature as $SO(2)\times SO(2)\subset SO(5)$ (or $SO(2)\times SO(3)\subset SO(5)$ for the $\CN=2$ preserving case), and then use relations between the Pontryagin classes of real bundles and the Chern roots of their complexified covers---see Appendix \ref{sec:anomalypolynomials} for relevant formulae.

The curvature for the normal bundle $NW_6$ has two contributions: one from the Riemann surface, and the other from the four-dimensional spacetime $M^{1,3}$.  Then, the roots of the normal bundle, which we'll denote as $n_\pm$, can be written as
\begin{equation}
n_\pm = \hat{t}_\pm+  2c_1\left(U(1)_\pm\right),
\end{equation} where $c_1\left(U(1)_\pm \right)$ is the first Chern class of the $U(1)_\pm$ symmetries of class $\mathcal{S}$, and $\hat{t}_\pm$ is the contribution of the curvature of the $SO(2)_\pm$ bundles over the Riemann surface.  The Calabi-Yau condition, or the topological twist, restricts the $\hat{t}_\pm$ as 
\begin{equation}
\hat{t}_+ +\hat{t}_- + \hat{t} =0, \qquad \int_{\Sigma_{g,n}} \hat{t} = \chi \left(\Sigma_{g,n} \right). \label{eq:background}
\end{equation} In these expressions, $\hat{t}$ is the curvature of the tangent bundle of the Riemann surface.  It will be useful to introduce the connection one-forms $A_\pm$ and their curvatures, $F_\pm$, as
\begin{equation}
\hat{t}_\pm = \frac{1}{2\pi} d A_\pm \equiv \frac{1}{2\pi} F_\pm.
\end{equation}  

When there are punctures on the Riemann surface, the curvatures $F_\pm$ can pick up sources localized at the punctures. This will imply that the region near the puncture must be replaced by a new geometry, whose properties are fixed by the choice of sources for $F_\pm$.  We denote by ${M}_6$ the total space of the $SO(5)$ bundle over the Riemann surface, and $X_6$ the local geometry near the puncture. Part of the story of understanding punctures is understanding the possible choices of the local geometry $X_6$.

If one of the contributions $\hat{t}_\pm$ is trivial, the compactification preserves eight supercharges and the four-dimensional quantum field theory preserves $\mathcal{N}=2$ supersymmetry.  Without loss of generality, we choose our $\mathcal{N}=2$ limit to be $\hat{t}_- =0$, in which case the $U(1)_-$ symmetry enhances to an $SU(2)_-$ R-symmetry.  In this limit, we have the following parametrization:
\begin{equation}
\hat{t}_+ = \frac{d A}{2\pi}  \equiv \frac{F}{2\pi} , \qquad n_+ = \frac{F}{2\pi}+ 2 c_1^+, \qquad \hat{t}_-=0, \qquad n_-^2 = -4 c_2^-.  \label{eq:N2con}
\end{equation} Here we have dropped the $(+)$ subscript on $F$ since $\hat{t}_-$ is trivial, and we are utilizing a shorthand notation
\begin{equation}
c_1^+ = c_1(U(1)_+), \qquad c_2^- = c_2(SU(2)_-).
\end{equation}

In this paper, we aim to argue for the structure of class $\mathcal{S}$ anomalies as presented above.  For conceptual clarity, we restrict to systems with eight supercharges.  The inflow analysis for systems with four supercharges can be discussed in the same way, however the reduction is reasonably more involved.

\subsection*{Angular forms} 

Now, we explain how to construct the angular form $e_4$ that appears in the Bianchi identity \eqref{eq:thom} to reflect the restricted $U(1)\times SU(2)$ isometry manifest on the four-sphere transverse to the M5-branes.

When the M5-branes are curved, the normal bundle is reduced.  The magnetic source for $G_4$ must be suitably modified to reflect this.  The source can be written alla \eqref{eq:thom} as
\begin{equation}
d G_4 = d\rho(r) \wedge d \Omega_4 (\widetilde{S}^4)
\end{equation} where $d\rho(r)$ is the smoothing of $\delta^5(r)r^4 dr$.  Here, the angular form $d\Omega_4(\widetilde{S}^4)$ is for a four-sphere $\widetilde{S}^4$ that is not maximally symmetric.  

This volume form depends on the normal bundle.  
If the twist preserves eight supercharges, the normal bundle of the branes has a $U(1)\times SU(2)$ structure group and therefore only a $U(1)\times SU(2)$ isometry is manifest on the four-sphere.  The connection of the $U(1)$ has a nontrivial component over the Riemann surface, while the connection of the $SU(2)$ over the surface is trivial in order to preserve the $SU(2)$ symmetry.  A metric over the four-sphere can be chosen as
\begin{equation}
ds^2(\widetilde{S}^4)= \frac{d\mu^2}{1-\mu^2} + (1-\mu^2) d\phi^2 + \mu^2 ds^2(S^2_\Omega),\label{eq:metric}
\end{equation}  with $\mu$ the interval $[0,1]$. The gauge invariant volume form is then
\begin{equation}
D \Omega_4 = \frac{1}{V_4} \mu^2 d\mu \wedge D\phi  \wedge D \Omega_2, \qquad D\phi \equiv d\phi -A_\phi - A,  \label{eq:e4N2}
\end{equation} where $V_4$ is the area of the four-sphere, $V_4=8\pi^2/3$, and $D\Omega_2$ is the gauge-invariant volume form of the round two-sphere $S_\Omega^2$ given as
\begin{equation}
 D\Omega_2 = \frac{1}{2} \epsilon_{abc} D \hat{y}^a \wedge D \hat{y}^b \hat{y}^c, \qquad D\hat{y}^a = d\hat{y}^a - A^{ab}\hat{y}^b, \qquad \sum_{a=1}^3 (\hat{y}^a)^2 =1.
\end{equation}  $A^{ab}$ is the connection for an $SO(3)$ bundle over the branes, with corresponding field strength 
\ba{
F^{ab} = d A^{ab} - A^{ac} \wedge A^{cb}.
}
The connection $A_\phi$ is the contribution over the flat four-dimensional space, and $A$ is the contribution over the Riemann surface. The corresponding curvatures are
\begin{equation}
dA = F, \qquad d A_\phi = F_\phi.  
\end{equation} 

While the angular form \eqref{eq:e4N2} is gauge invariant, it is not closed.  The most general closed and gauge invariant angular form can be written as 
\begin{align}
E_4 = \frac{1}{3V_4} d \left[ \mu D\phi \wedge \left(\mu^2D\Omega_2 - h(\mu) F_2^\Omega \right) + \left(a_\phi A_\phi + a_s A\right) \wedge e_2^\Omega\right]  \label{eq:genE4}
\end{align} where we have introduced an arbitrary function $h(\mu)$ and arbitrary constants $(a_\phi, a_s)$.  The $SO(3)$ forms are given as
\begin{equation}
F^\Omega_2 = \frac{1}{2} \epsilon_{abc} F^{ab} \hat{y}^c, \qquad e_2^\Omega = D\Omega_2 -F_2^\Omega, \qquad d(D\Omega_2 ) =  F^\Omega_3 =\frac{1}{2} \epsilon_{abc} F^{ab} \wedge D\hat{y}^c.  
\end{equation} One choice of $h(\mu)$ and the $a$'s corresponds to taking the $SO(5)$ gauge invariant angular form in \eqref{eq:e4so5} and reducing it so that only an $SO(2)\times SO(3)$ is manifest.  This choice corresponds to $h(\mu)=1$ and $a_\phi =a_s =0$.  

\subsection{Components of inflow with punctures}

Here, we give an overview of the new elements needed to account for the puncture data in the inflow procedure. We argue that in the inflow procedure, the integrand will split into two independent contributions. We finish with a computation of the bulk contributions to the Class $\CS$ anomalies. 

\subsubsection*{Sources for connection forms}

The curvature form is not well-defined at the locations of punctures on the Riemann surface.  The connection one-forms over the Riemann surface $A_\pm$ are singular at the punctures, and suitable boundary conditions are needed.  Motivated by the work in gravity \cite{Gaiotto:2009gz,Bah:2015fwa}, we allow for explicit sources for the connection localized at the punctures. The class of examples we will consider here are cases where the sources are monopoles. We propose that in order to account for punctures in the inflow computation we should add magnetic sources for the curvature forms $F_\pm$,  as
\begin{equation}
d F_\pm = 2\pi \sum_{\alpha=1}^n \delta(\vec{x} -\vec{x}_\alpha ) df_\alpha^\pm \wedge dx^1 \wedge dx^2.
\end{equation} The $(x^1,x^2)$ are coordinates on the Riemann surface and $\vec{x}_\alpha$ is the location of a puncture with label $\alpha$.  The functions $f_\alpha^\pm$ depend on the transverse coordinates and encode the boundary data for the connection one-forms.  The allowed choices of $f_\alpha^\pm$ are constrained by supersymmetry.  


We restrict now to cases with one puncture, $\alpha=1$, and reductions that preserve eight supercharges.  We can turn on multiple monopole sources for $F$ along the $\mu$ interval,   
\begin{equation}
d F = 2\pi \delta(\vec{x} -\vec{x}_1 )  \sum_{a} d f^a(\mu) \wedge dx^1 \wedge dx^2, \qquad df^a =   \hat{k}^a \delta\left(\mu -\mu^a\right)d\mu,\label{eq:sourcea}
\end{equation} where $a$ labels the different monopoles, and the $\hat{k}^a$ are constants related to the monopole charge that should be determined by flux quantization in the near-puncture geometry.

The local geometry should be determined by a space with smoothed out monopole sources. In particular, we have the decomposition
		\ba{\bs{
	\text{bulk:}&\qquad {M}_6 = S^2_\Omega\times S^1_\phi   \times [\mu]\times \Sigma_{g,1}\ ,\\
	\text{near-puncture:}& \qquad X_6 = S^2_\Omega \times X_4\  .
	}\label{eq:bundlesplit}}
The geometry $X_4$ is fixed by the choice of sources for $F$. In particular, the correct bundle structure for $X_4$ is not necessarily the one inherited from the bulk, as is indeed the case for the holographic dual to the punctured system  \cite{Gaiotto:2009gz}.  Quantization conditions for $F$ will depend on the properties of $X_4$.

Each source in \eqref{eq:sourcea} corresponds to a monopole located at $\left(\vec{x} =\vec{x}_1, \mu = \mu^a\right)$.  In M-theory, this source is a co-dimension three object whose world-volume we denote as $W_8$.  The tangent bundle of M-theory near the source decomposes as
\begin{equation}
TM_{11}|_{W_8} = TW_8 \oplus NW_8, \label{eq:splitting}
\end{equation} where $TW_8$ is the curvature bundle on the source and $NW_8$ is an $SO(3)$ normal bundle.  The diffeomorphisms of M-theory induce an $SO(3)$ gauge symmetry on the world-volume of the source.  The background geometry where the source lives splits the $\mu$ direction from the $(x_1, x_2)$ directions, and therefore only a $U(1)$ subgroup of this $SO(3)$ gauge symmetry group is preserved.    

This description of the sources is consistent with the picture in the holographic duals \cite{Gaiotto:2009gz,Bah:2015fwa}. In these solutions, there are additional M5-branes that are localized at the punctures.  These branes are extended along a direction normal to the Riemann surface and end at monopole sources of a $U(1)$ connection of an $S^1$ bundle over the  surface, which here corresponds to the $A$ connection. The locations of the monopole sources along the $\mu$ interval are denoted here as $\mu^a$, and should be quantized\footnote{The $\mu$ interval is the $y$ interval in the backreacted systems in \cite{Gaiotto:2009gz}.}.

More general $f^a$  in \eqref{eq:sourcea} could be obtained by smearing the delta function source. Here, we will not discuss the smoothing out of the source, since this requires a better understanding of the geometry $X_4$. Schematically, one would write an expression 
\begin{align}
F = dA + \sum_a g_2^a.  \label{eq:Fg}
\end{align} 
The first term $dA$ is the flux associated to the holonomy of the Riemann surface, which contributes to the Euler characteristic. In particular, the background curvature of the tangent bundle of the surface $\hat{t}$, discussed around \eqref{eq:background}, still satisfies 
\begin{equation}
dA = -2\pi \hat{t}, \qquad \int_{\Sigma_{g,1}} dA = -2\pi \chi(\Sigma_{g,1} ).\label{eq:useful1}
\end{equation} 
The second set of terms in \eqref{eq:Fg} will depend explicitly on data at the monopoles. 
In the near-puncture region, $F$ satisfies the boundary condition that $dA\to 0$; i.e. the connection is flat.  This can be seen by looking at the background metric near the puncture \cite{Bah:2015fwa}.

\subsubsection*{Consistent sources for $G_4$}

 Sources for the curvatures $F=dA$ induce sources for the four-form flux $G_4$.  Since the gauge invariant angular form in \eqref{eq:e4N2} has an explicit dependence on $F$, $G_4$ cannot be closed in the presence of sources for $F$.  It needs to be further improved.  

To understand the sources induced for $G_4$, we temporarily turn off the connections on the four-dimensional space.  Then, the closed magnetic sources for $G_4$ in the presence of $N$ M5-branes are
\begin{align}
d G_4 &=\frac{1}{V_4} d\rho(r) \wedge \left(\mu^2 d\mu \wedge \left(d \phi -A\right) + \frac{1}{3} \left(  a_s-\mu^3 \right) F \right) \wedge d\Omega_2 +\frac{1}{V_4}   K_3 \wedge d\Omega_2, \label{eq:sourceF}\\
d K_3 &= \frac{1}{3} \left( a_s - \mu^3\right) d\rho(r) \wedge dF.  
\end{align}  The $K_3$ term is needed to close the source term for $G_4$ in presence of the monopoles.  We observe that consistency of the sources requires the M5-branes wrapped on the Riemann surface to branch off at the punctures.   This is consistent with the probe analysis for punctures in holography \cite{Gaiotto:2009gz,Bah:2013wda}. 

The crucial next step is to construct ${E}_4$ as in  \eqref{eq:genE4} in the near-puncture region. The construction of the global angular form $E_4$ depends intimately on the structure of the local geometry $X_6$. Because $E_4$ depends on the curvature $F$, the inclusion of monopole source terms for the curvature $F$ will give new terms in $E_4$. Denote by $\widetilde{E}_4$ the improved, closed, and globally well-defined ${E}_4$ in the near-puncture region, which reduces to to the bulk $E_4$  \eqref{eq:genE4}  away from the punctures.
We can then write the most general closed, gauge invariant, and global source for $G_4$ in the near-puncture region as
\begin{equation}
d G_4 =  d \rho(r) \wedge \widetilde{E}_4. \label{eq:e4punct}
\end{equation}  Our convention is $\rho(0)=-N$ and $\rho(r\to\infty)=0$.

In addition to the condition that $G_4$ integrates over the $S^4$-bundle to $N$, $G_4$ will satisfy other quantization conditions in $X_6$. For example, the integral of $G_4$ on the 4-cycle consisting of the $S^2_\Omega$ times the 2-cycle connecting monopoles should be quantized, as well as the integral of $G_4$ on the 4-cycle consisting of the $S^2_\Omega$ times the sphere surrounding the monopole (discussed around \eqref{eq:splitting}). The correct quantization conditions will depend on a more careful understanding of the local geometry $X_4$ near the puncture region, which we leave to future work \cite{Bah:2018bn}.

\subsubsection*{Variation of the action}

The variation of the action has two terms, given as 
	\ba{
	\frac{\delta S}{2\pi}  =- \frac{1}{6} \delta  \int_{\widetilde{M}_{11}} {C}_3 \wedge {G}_4 \wedge {G}_4 {-} \delta \int_{\widetilde{M}_{11}} {C}_3 \wedge {I}_8^{\text{inf}}.\label{eq:totvar}
	}
We've excised small regions around the M5-branes $(r<\epsilon)$ from $M_{11}$ to obtain $\widetilde{M}_{11}$. The variation of the action is computed by integrating over $\widetilde{M}_{11}$, and then taking $\epsilon$ to zero. 

In the region near $r=\epsilon$, we split the eleven-dimensional manifold as $\widetilde{M}_{11}=[r]\times M^{1,3}\times {M}_{6}$, where the six-dimensional part is the total space of the $S^4$ bundle over the Riemann surface, as given in \eqref{eq:bundlesplit}. We then excise a small disk around the puncture on the Riemann surface to obtain a smooth space $\widetilde{M}_6$. The excised portion is replaced with the local geometry $X_6=S^2_\Omega\times X_4$ in  \eqref{eq:bundlesplit}.

In the near-puncture region $G_4$ is given in terms of $\widetilde{E}_4$, as discussed around \eqref{eq:e4punct}. Depending on the form of $X_4$, $\widetilde{E}_4$ will have various terms that localize at the puncture.   The integrals in \eqref{eq:totvar} will only pick up contributions near the sources, and the integral will localize over $X_6$. Away from the punctures, $\widetilde{E}_4$ must reduce to the bulk $E_4$ given in  \eqref{eq:genE4}, since the $G_4$ flux must be smooth everywhere. Then, the integrands that saturate $\widetilde{M}_6$ will be given in terms of $E_4$.

Due to these considerations, the variation will split into two terms as
	\ba{
	\frac{\delta S}{2\pi} = \int_{M^{1,3}\times \widetilde{M}_6} \CI_{10}^{(1)}+  \int_{M^{1,3}\times X_6} \CI_{10}^{(1)}\ , \label{eq:decomp0}
	}
where $\CI_{10}^{(1)}$ is a ten-form determined from \eqref{eq:totvar}. 
This translates into an anomaly polynomial 
\ba{
I_6^{\CS} &= \int_{\widetilde{M}_6} \CI_{10}^{(1)}  + \int_{X_6} \CI_{10}^{(1)} \label{eq:decomp1} \\
&=I_6(\Sigma_{g,1}) + I_6(P). \label{eq:decomp2}
}
These expressions validate the general expectation of the structure of anomalies of class $\mathcal{S}$ theories as described in \eqref{eq:cs}.

Let us consider the two terms in \eqref{eq:totvar} in turn, and compute the bulk contributions to the anomaly polynomial (the first terms in each of \eqref{eq:decomp0},  \eqref{eq:decomp1},\eqref{eq:decomp2}). 
We will not explicitly compute the contribution due to the punctures, since this requires an understanding of the local geometry $X_6$, and the angular form $\widetilde{E}_4$ in that region. 

First, consider the variation of the Chern-Simons term. The variation of the action away from the punctures gives
\begin{align}
\frac{\delta S_{CS}}{2\pi} &=\frac{1}{6} \int_{\widetilde{M}_{11}} d \left(\rho {E}_2^{(1)} \right) \wedge d \left(\rho {E}_3^{(0)} \right) \wedge d \left(\rho {E}_3^{(0)} \right) \\
&= \frac{1}{6} \int_{[r]} d\rho^3 \left[  \int_{M^{1,3}\times \widetilde{M}_6} {E}_2^{(1)} \wedge {E}_4 \wedge  {E}_4    \right]. \label{eq:varcs}
\end{align} In evaluating the variation, we dropped terms involving $C_3$.  The integrand factorizes in its $r$ dependence and therefore we can pull out the overall $\rho$ dependence. 
 Since $\rho$ vanishes as $r\to \infty$, the only contribution comes from $r=\epsilon$ where we have
\begin{equation}
\int_{[r]} d\rho^3 = -N^3. \label{eq:rho}
\end{equation}
The bulk term in \eqref{eq:decomp0} then yields the descent of the eight-form anomaly polynomial,
\begin{equation}
\left[I_{CS,8}^{\text{bulk}} \right]^{(1)}  = \frac{N^3}{6}\int_{S^1_\phi\times S^2_\Omega\times[\mu]}{E}_2^{(1)}\wedge {E}_4\wedge {E}_4.\label{eq:csbulkf}
\end{equation}
In general, the anomaly will depend on the choice of the function $h(\mu)$ in \eqref{eq:genE4}.  In this paper we will not analyze the general case, and will instead fix them to match onto the reduction of the $SO(5)$ bundle to $SO(2)\times SO(3)$, with $h(\mu)=1$ and $a_\phi = a_s=0$.

We reconstruct $I_{CS,8}^{\text{bulk}}$ from \eqref{eq:csbulkf} as\footnote{The story will be similar for general $h(\mu)$ and $a_\phi,a_s$, When we compute \eqref{eq:csbulkf}, we will find that the only terms that survive the integral over $[\mu]\times \Sigma_{g,1}$ will be proportional to $dA$. Then, $n_+$ is independent of $\mu$ and can be pulled out of the integral. The effect will be to simply multiply the answer for $I_{CS,6}^{\text{bulk}}$ by a function of $a_\phi$ and $a_s$. These constants can be fixed by regularity and matching conditions.
We do not consider this more general case here.}
	\ba{
	I_{CS,8}^{\text{bulk}} =\frac{N^3}{24} {n_+^2n_-^2}=  \frac{N^3}{24} {p_2(NW_6)}.\label{eq:csbulka}
	}
The bulk anomaly contribution to the anomaly polynomial of the four-dimensional theory can then be computed from \eqref{eq:csbulka} as
\begin{align}
I_{CS,6}^{\text{bulk}} &= \int_{\Sigma_{g,1}} I_{CS,8}^{\text{bulk}} = - \frac{2N^3}{3} \int \frac{dA}{2\pi}\wedge c_1^+ c_2^- = \frac{2N^3}{3} \chi(\Sigma_{g,1}) \;   c_1^+ c_2^-. \label{eq:IbulkCS}
\end{align}
Recall that the relation of the roots of the normal bundle $n_\pm$ to the Chern roots of the $U(1)_+ \times SU(2)_-$ is given in \eqref{eq:N2con}.

Next we evaluate the variation of the linear term away from the punctures.  For this, we need to first reduce $I_8^{\text{inf}}$ in \eqref{eq:iinf} and then restrict to the case with eight supercharges.  Under the decomposition of the curvature bundle, we have
\begin{align}
p_1(R) &= p_1(T^4) + \hat{t}^2 + n_+^2 + n_-^2 \label{eq:p1}\\
p_2(R) &= p_2(T^4) +p_1(T^4) \left( \hat{t}^2 + n_+^2 + n_-^2 \right) + \hat{t}^2 (n_+^2 + n_-^2) + n_+^2 n_-^2.\label{eq:p2}
\end{align} The relation of the roots of the normal bundle, $n_\pm$, to the Chern roots of the $U(1)_+ \times SU(2)_-$ is given in \eqref{eq:N2con}.  The curvature of the Riemann surface is given as $\hat{t}$, which satisfies \eqref{eq:useful1}.

The variation of the linear term goes as
\begin{align}
- \delta \int {C}_3 \wedge  {I}_8^{\text{inf}} &=- \delta \int {G}_4 \wedge  {I}_7^{\text{inf}(0)} = - \int {G}_4 \wedge d  {I}_6^{\text{inf}(1)} = \int d\left(\rho {E}_3^{(0)} \right) \wedge d  {I}_6^{\text{inf}(1)} \\
&= \int_{[r]} d\rho \left[ \int_{M^{1,3} \times \widetilde{M}_6} {E}_4 \wedge  {I}_6^{\text{inf}(1)}  \right].
\end{align}

We reduce the eight-form while only keeping terms that can be non-trivial in the action, leading to
\begin{align}
 {I}_8^{\text{inf}} &\supset \frac{1}{192}  \left(\frac{{F}}{2\pi} + 2c_1^+ \right)^4 -\frac{1}{96} \left( p_1(T^4) - 4 c_2^- \right)\wedge \left(\frac{{F}}{2\pi} + 2c_1^+ \right)^2.
\end{align} 
Therefore, we have
\begin{equation}
\left[I_{L,8}^{\text{bulk}} \right]^{(1)} = N \int_{S^1_\phi\times S^2_\Omega\times[\mu]}{E}_4 \wedge  {I}_6^{\text{inf}(1)}.\label{eq:lbulkf}
\end{equation}
 Since the polynomial $I_8^{\text{inf}}$ has no dependence on the angular coordinates, it follows that the only contribution from $E_4$ is actually the volume form of the transverse four-sphere.  This expression is independent of any choice for $h(\mu)$ in \eqref{eq:genE4}.  We can then reconstruct the bulk contribution simply as 
	\ba{
	I_{L,8}^{\text{bulk}} = N I_8^{\text{inf}}.
	}
Integrating over the Riemann surface, we compute the contribution to the six-form anomaly polynomial of the class $\CS$ theory as
\ba{
I_{L,6}^{\text{bulk}} &= \int_{\Sigma_{g,1}} I_{L,8}^{\text{bulk}} = -\frac{\chi(\Sigma_{g,1}) }{2} N \left( \frac{(c_1^+)^3}{3}  - \frac{c_1^+ p_1(T^4) }{12} + \frac{c_1^+ c_2^-}{3}   \right).  \label{eq:IbulkL}
}

Together, the bulk terms  computed in \eqref{eq:IbulkCS} and \eqref{eq:IbulkL} yield
	\begin{equation}
	I_6(\Sigma_{g,n}) = -\frac{\chi(\Sigma_{g,n})}{2} \left[ N  \left( \frac{(c_1^+)^3}{3}-	\frac{c_1^+p_1(T^4)}{12} + \frac{c_1^+ c_2^-}{3}\right)- \frac{4}{3} N^3 c_1^+ c_2^- \right]\ , 	\label{eq:answer1}
	\end{equation} 
where we've generalized to the case of $n$ punctures. Indeed, $I_6(\Sigma_{g,n})$ is of the form described in \eqref{eq:integrate1} where we integrate the polynomial from the M5-branes given in \eqref{eq:I8flat}.  The difference between the result of this computation and the anomaly polynomial of the $A_{N-1}$ $(2,0)$ theory is due to an overall free  tensor multiplet that decouples from the dynamics of the M5-branes.

\section{Conclusions}

\subsection{Summary}

In Section \ref{sec:n1}, we studied the anomalies of the $\CN=1$ four-dimensional theories that derive from M5-branes wrapped on a punctured Riemann surface. When the bulk preserves $\CN=1$, the R-symmetry locally at a puncture can preserve $\CN=2$ supersymmetry with a local twist of the R-symmetry generators labeled by integers $(p,q)$. We derived the anomalies of these $(p,q)$-labeled punctures, and discussed an illuminating parameterization of the anomalies in terms of an $\CN=1$ version of an effective number of vector and hypermultiplets.

In Sections \ref{sec:inflow1} and \ref{sec:inflow2}, we turned to the problem of computing the anomalies of the $\CN=2$ theories of class $\CS$ from inflow for M5-branes wrapped on a punctured Riemann surface. In our analysis, we motivated the addition of monopole sources at the locations of the punctures for the connection form on the Riemann surface. These appear as delta functions on the right-hand-side of the Bianchi identity for the associated curvature $F$, which then have to be appropriately smoothed.

The M5-branes magnetically source the M-theory flux $G_4$. The sources for the connection form on the surface induce additional sources for $G_4$.  Compatibility conditions between these sources require additional M5-branes that intersect the original ones at the punctures and end on the monopole sources. This is consistent with the picture in AdS/CFT.  When the branes are backreacted, there is an $AdS_5$ spacetime that emerges in the near-horizon limit of the branes.  The connection forms in a Ricci flat background pick up such monopole sources in the work of \cite{Gaiotto:2009gz,Bah:2015fwa}.  

We outlined a procedure to derive the anomaly contributions from the additional degrees of freedom at the punctures.  In order to proceed with the computation of the puncture contribution, we must address the following questions:

\begin{itemize}

\item The form of the local geometry $X_6$ in the near-puncture region.

\item How to fix the precise form of the angular form $\widetilde{E}_4$, as well as the free parameters $a_\phi,a_s$. The possible choices are intimately related to regularity conditions on the flux in various limits.

\item How the parameters $\mu^a$ encode the data of the punctures in field theory. In particular, for the case of regular punctures these should be associated to the data of the Young diagram that corresponds to the flavor symmetry preserved in the CFT. One can hope to extend this analysis to the case of irregular punctures. 

\item The relation of the $U(1)$ symmetries at the monopole locations to the global symmetries of the CFT.

\item What are the decoupled modes of the system? In particular, the inflow result includes modes that decouple with respect to the low energy CFT.

\end{itemize}
	
These issues will be addressed in the upcoming \cite{Bah:2018bn}.

\subsection{Outlook}

In this paper we have focused on four-dimensional field theories that preserve $\CN=2$ supersymmetry. The generalization to the $\CN=1$ theories will follow the steps we've laid out here, but with some interesting additional complications. In particular,  in the $\CN=1$ case the normal bundle to the M5-branes decomposes as $SO(2)_+\times SO(2)_- \subset SO(5)$. The field strengths for each $SO(2)_\pm$ will have sources at the punctures, whose profile in the normal directions can be more involved. We leave the analysis of the $\CN=1$ story in future work.

The $\CN=1$ theories of Class $\CS$ are even more rich than their $\CN=2$ counterparts. One feature is that different kinds of punctures can be present in the $\CN=1$ class $\CS$  construction. One example are the class of $(p,q)$-labeled punctured we discussed in Section \ref{sec:n1}. It would be interesting to understand these anomalies from an inflow analysis. More generally, the landscape of $\CN=1$ preserving punctures in these geometries is much less understood than their $\CN=2$ counterparts, and would be interesting to study further.

In the region near the puncture, the M-theory system can be reduced to Type IIA string theory, and the degrees of freedom at the puncture are associated to the intersection of D4/D6 branes \cite{Witten:1997sc}. From this perspective, the contributions from the punctures should be related to I-branes, as discussed in \cite{Green:1996dd}. Such intersections are also related to D6/D8 brane intersections, which appear in the classification of (1,0) theories \cite{Gaiotto:2014lca}. It would be interesting to explore these connections in the future. 

Throughout this paper, we have only discussed theories which have their origin from the (2,0) theories in six dimensions. One could also consider starting from theories with less supersymmetry, such as six-dimensional (1,0) SCFTs (with a recently proposed classification in \cite{Heckman:2013pva,Heckman:2015bfa,Bhardwaj:2015xxa}). These are far more numerous, and their compactifications are less understood than their (2,0) counterparts. One large class of such constructions---dubbed class $\mc{S}_k$---involve $N$ M5-branes on an $A_{k-1}$ singularity of M-theory compactified on a punctured Riemann surface \cite{Gaiotto:2015usa,Hanany:2015pfa,Razamat:2016dpl,Bah:2017gph}. It would be very interesting to extend the inflow analysis we considered here in the class $\CS$ context to these theories. 

As a specific example, 't Hooft anomalies for four-dimensional theories that result from compactifications of the six-dimensional E-string theory on a punctured Riemann surface were computed in \cite{Kim:2017toz}. There, the contributions of the punctures to the anomalies were obtained by studying boundary conditions of the E-string theory at the punctures, and adding up the anomalies from chiral fields living on the boundary. It would be interesting to understand these contributions from the perspective we've advocated here.

\section*{Acknowledgements}

We thank Ken Intriligator for many discussions throughout this project. We are grateful to Federico Bonetti, Ken Intriligator, and Yuji Tachikawa for discussion and for looking over an earlier version of the manuscript.  We additionally thank Nikolay Bobev, Thomas Dumitrescu, Kazunobu Maruyoshi, Ruben Minasian, and Jaewon Song for discussion. EN is supported in part by DOE grant DE-SC0009919, and a UC President's Dissertation Year Fellowship.

We dedicate the paper to IB's father who recently passed away.  I, Ibrahima, am very grateful for all of his hard work and sacrifices which has lead to all that I am; you will be missed.

\appendix

\section{SCFT Facts}
\label{sec:scfts}

\subsection{$\CN=1$}

Conformal field theories in even spacetime dimensions have Weyl anomalies. The conformal anomaly of the trace of the four-dimensional energy momentum tensor on a curved background is
	\ba{
	\langle T_\mu^\mu \rangle &= -\frac{1}{16\pi^2} \big[ {a}(\text{Euler})-{c} (\text{Weyl})^2\big],
	}
where
	\ba{
	 (\text{Weyl})^2 &= (R_{\mu\nu\rho\sigma})^2 - 2 (R_{\mu\nu})^2 + \frac{1}{3} R^2,\quad
	 (\text{Euler}) = (R_{\mu\nu\rho\sigma})^2 - 4 (R_{\mu\nu})^2 + R^2.
	}
The coefficients $a$ and $c$ are the central charges of the CFT. For an SCFT preserving $\mc{N}=1$ supersymmetry, they are related to the 't Hooft anomalies of the superconformal $U(1)_R$ symmetry by the superconformal algebra \cite{Anselmi:1997am},
	\ba{
	a = \frac{3}{32}\left( 3 \tr R^3_{\mc{N}=1} - \tr R_{\mc{N}=1}\right),\quad c = \frac{1}{32} \left( 9 \tr R^3_{\mc{N}=1} - 5 \tr R_{\mc{N}=1}\right).
	}
If the theory has a flavor symmetry $G$ with generators $T^a$, the flavor central charge is defined as
	\ba{
	k_G\delta^{ab}=-3\tr R_{\CN=1}T^aT^b.
	}

\subsection{$\CN=2$}

An $\mc{N}=2$ SCFT has an R-symmetry $U(1)_R\times SU(2)_R$, with generators $R_{\mc{N}=2}$ and $I^a$ $(a=1,2,3)$ respectively. We use a basis for the Cartan subalgebra of the R-symmetry labeled by $(R_{\mc{N}=2},I^3)$. The R-charge assignment for free $\mc{N}=2$ vector multiplets and hypermultiplets is 
	\begin{table}[h!]
	\centering
	\begin{tabular}{c|ccc}
{$R_{\mc{N}=2}\ \backslash\  I^3$}
 & $\frac{1}{2}$ & 0 & $-\frac{1}{2}$ \\ \hline
	 0	& & $A_\mu$ & \\
	 1 	& $\lambda_\alpha$ & & $ \lambda_\alpha'$ \\
	 2 	& & $\phi$ & 
	\end{tabular} \qquad
	\begin{tabular}{c|ccc}
{$R_{\mc{N}=2}\ \backslash\  I^3$} 
 & $\frac{1}{2}$ & 0 & $-\frac{1}{2}$ \\ \hline
	 -1	& & $\psi_\alpha$ & \\
	 0 	& $Q$ & & $ \widetilde{Q}^\dagger$ \\
	 1 	& & $\overline{\widetilde{\psi}}_{\dot{\alpha}}$ & 
	\end{tabular}
	\end{table} 
	
With this charge assignment, the nonzero anomaly coefficients for an $\mc{N}=2$ vector are $\tr R_{\mc{N}=2}=\tr R_{\mc{N}=2}^3=2,$ and $\tr R_{\mc{N}=2} (I^3)^2 = \frac{1}{2}$. The nonzero coefficients for an $\mc{N}=2$ hypermultiplet are $\tr R_{\mc{N}=2}= \tr R_{\mc{N}=2}^3=-2$.

The central charges $a$ and $c$ are related to the anomaly coefficients as \cite{Kuzenko:1999pi}
	\ba{
	\tr R_{\mc{N}=2}^3 = \tr R_{\mc{N}=2} = 48(a-c),\qquad \tr R_{\mc{N}=2} I^aI^b=2\delta^{ab}(2a-c).
	}
The flavor central charge $k_G$ for a global symmetry $G$ with generators $T^a$ is
	\ba{
	k_G \delta^{ab}=-2\tr R_{\CN=2} T^aT^b.\label{eq:n2kcoef}
	}
Using $n_v$ and $n_h$ to represent the number of free vector multiplets and free hypermultiplets, the central charges of an $\mc{N}=2$ superconformal theory can be written
	\ba{
	a = \frac{1}{24} \left( n_h + 5 n_v\right),\qquad c = \frac{1}{12} \left( n_h + 2 n_v\right), \label{eq:charges}
	}
or in other words,
	\ba{
	\tr R_{\mc{N}=2}^3 = \tr R_{\mc{N}=2} = 2(n_v-n_h),\qquad \tr R_{\mc{N}=2} I^aI^b=\delta^{ab}\frac{n_v}{2}.\label{eq:n2coef}
	}
	
We can fix an $\mc{N}=1$ subalgebra in the $\mc{N}=2$ algebra, such that the $\mc{N}=1$ R-symmetry is given by
	\ba{
	R_{\CN=1} = \frac{1}{3} R_{\CN=2} + \frac{4}{3} I_3.\label{eq:n1r}
	} 
With this choice, the linear combination
	$J=R_{\mc{N}=2} - 2I_3$
commutes with this $\mc{N}=1$ subalgebra, and thus is a flavor symmetry from the $\mc{N}=1$ point of view. (\ref{eq:n1r}) is the unique $\CN=1$ R-symmetry that has the properties of a superconformal $U(1)_R$ when the theory has enhanced $\CN=2$ supersymmetry.

\section{Anomaly Polynomials and Characteristic Classes} \label{sec:anomalypolynomials}

As reviewed in the main text, anomalies are encoded in a $(d+2)$-form anomaly polynomial $I_{d+2}$ that is related to the anomalous variation of the quantum effective action as
	\ba{
	\delta {S}_{\text{eff}} = 2\pi \int_{M_d} I_{d}^{(1)},
	}
where 
	\ba{
	I_{d+2}=dI_{d+1}^{(0)},\qquad \delta I_{d+1}^{(0)} = dI_{d}^{(1)}.
	}
Anomalies for chiral fields in even $d=2n$ dimensions are related to index theorems in two higher dimensions \cite{AlvarezGaume:1983ig}. For example, the Atiyah-Singer index theorem for a chiral spin-1/2 fermion in $d+2$ dimensions relates the index density of the Dirac operator to characteristic classes of the curvatures, which in turn are related to the $(d+2)$-form anomaly polynomial as
	\ba{
	I_{d+2} = \text{index}(i \slashed{D})=\big[ \hat{A}(R)\text{ch}(F)\big]_{d+2}. \label{eq:index}
	}
The $(d+2)$ subscript in \eqref{eq:index} instructs us to extract the $(d+2)$-form contribution in the expansion of the curvatures. $\text{ch}(F)$ is the Chern character, defined for a complex bundle in terms of the corresponding field strength $F$ as
	\ba{
	\text{ch}(F)=\tr_{\bf r} e^{iF/(2\pi)}= \text{dim}({\bf r}) + c_1(F) + \frac{1}{2}(c_1(F)^2 - 2c_2(F)) + \dots \label{eq:chernclass}
	}
The Chern classes $c_k$ are $2k$-forms that are polynomials in $F$ of degree $k$. For reference, the first two Chern classes take the form 
	\ba{
	c_1(F) =\frac{i}{2\pi}\text{Tr} F,\quad c_2(F) = \frac{1}{2(2\pi)^2} \left[\text{Tr}(F^2)-(\text{Tr}F)^2\right]. 
	}
Our notation is such that if the Chern roots of an $SU(N)$-bundle are given by $\lambda_i$, $c_2(F)=-\frac{1}{2}\sum_i \lambda_i^2$.

$\hat{A}(R)$ is the $A$-roof genus, a function of the curvature $R$ of the spacetime tangent bundle with leading terms
	\ba{
	\hat{A}(R) = 1-\frac{1}{24}p_1(R) + \frac{7p_1(R)^2-4p_2(R)}{5760} +\dots
	}
The $p_k$ are the Pontryagin classes, $4k$-forms that are $2k$-order polynomials in $R$. For reference, the first two Pontryagin classes for a real vector bundle with curvature $R$ are
	\ba{
	p_1(R) &= -\frac{1}{2} \frac{1}{(2\pi)^2} \tr (R^2),\\
	p_2(R) &= \frac{1}{8} \frac{1}{(2\pi)^4}\left[\left(\tr (R^2)\right)^2- 2\tr (R^4)\right].
	}
For a real bundle with a complex cover, the Pontryagin classes can be related to the Chern classes. For an $SO(N)$ bundle $E$, $p_1(E)$ and $p_2(E)$ can be written in terms of the Chern roots $\lambda_i$ as 	
	\ba{
	p_1(E)=\sum_i \lambda_i^2, \qquad p_2(E)=\sum_{i<j} \lambda_i^2\lambda_j^2.
	}

Another useful set of identities relates the Pontryagin classes of a vector bundle which is the Whitney sum of two vector bundles, $E=E_1\oplus E_2$, to the Pontryagin classes of the constituent $E_{1,2}$ as
	\ba{
	p_1(E) &= p_1(E_1) + p_1(E_2)\\
	p_2(E) &= p_2(E_1) + p_2(E_2) + p_1(E_1)p_1(E_2). \label{eq:pont}
	}

From (\ref{eq:index}), it follows that the six-form anomaly polynomial for one four-dimensional Weyl fermion with $U(1)$ charge $q$ is
	\ba{
	I_6 = \big[\hat{A}(T^4)\text{ch}(qF)\big]_6 = \frac{q^3}{6} c_1(F)^3 - \frac{q}{24} c_1(F) p_1(T^4).
	}
Here, $F$ is the field strength of the $U(1)$ bundle, and $T^4$ is the spacetime tangent bundle. More generally, a four-dimensional theory with a $U(1)$ R-symmetry and anomaly coefficients $\tr R^3$ and $\tr R$ (n.b. that $R$ here does not refer to the curvature!) has the corresponding six-form anomaly polynomial:
	\ba{
	I_6 = \frac{\tr R^3}{6} c_1(F)^3 - \frac{\tr R}{24}c_1(F) p_1(T^4). \label{eq:geni}
	}
$F$ here is field strength of the $U(1)$ bundle coupled to the R-symmetry. $I_6$ is then related to the anomalous divergence of the R-symmetry current by the descent procedure.

\section{Anomalies for Regular $\CN=2$ Punctures} \label{sec:punctures}

A regular $\CN=2$ puncture is labeled by an embedding $\rho:\ \mathfrak{su}(2)\to \mathfrak{g}$. For $\mathfrak{g}=A_{N-1}$, the choice of $\rho$ is 1-to-1 with a partition of $N$, i.e. a Young diagram $Y$ with $N$ boxes. In this appendix, we review the contributions of punctures from the six-dimensional (2,0) $\mathfrak{g}=A_{N-1}$ theories compactified on a Riemann surface $\Sigma_{g,n}$ with genus $g$ and $n$ total punctures.

Let $Y$ have some number of columns of height $h_i$, and some number of rows of length $\ell_j$, corresponding to a partition of $N$  
	\ba{
	N = \sum_{i} h_i= \sum_{j} \ell_j.
	}
Let $n_i$ be the number of columns of height $h_i$ that appear in the sum. Then, the theory has an unbroken flavor symmetry
	\ba{
	G=S\left[\prod_iU(n_i)\right],
	}
which corresponds to the commutant of the embedding $\rho$. 

We also assign a pole structure to the puncture \cite{Gaiotto:2009we}, which can be read off of the Young diagram. Denote the pole structure by a set of $N$ integers $p_i$, $i=1,\dots, N$. Label each of the $N$ boxes in the Young diagram sequentially with a number from 1 to $N$, starting with $1$ in the upper left corner and increasing from left to right across a row. Then, $p_i=i-$(height of $i$th box). For instance, $p_1=1-1=0$ always.
	
For example, a ``maximal'' puncture is labeled by a Young diagram with $1$ row of length $\ell_1=N$, or alternatively, $N$ columns each of height $h_{1\leq j\leq N}=1$. This is commonly denoted $Y=[1,\dots,1]$. The unbroken flavor symmetry is $G=SU(N)$, and the pole structure is $p_i=i-1=(0,1,2,\dots,N-1)$. As another example, a ``minimal'' or ``simple'' puncture is labeled by a Young diagram with 1 row of length 2 and $N-2$ rows of length 1, or alternatively, 1 column of height $N-1$ and 1 column of height 1, denoted $Y=[N-1,1]$. The unbroken flavor symmetry is $G=S[U(1)\times U(1)]= U(1)$, and the pole structure is $p_i = (0,1,1,\dots,1)$. 

The effective number of vector multiplets that a regular puncture labeled by a Young diagram $Y$ contributes to the theory is \cite{Gaiotto:2009gz,Chacaltana:2010ks}
	\ba{\bs{
	n_v(P_Y)&=  -\frac{1}{2}\left( r_G + \frac{4}{3}d_Gh_G\right)+ \sum_{k=1}^N(2k-1)p_k \\
	}\label{eq:nvy}}
and the effective number of hypermultiplets is 
	\ba{\bs{
	n_h(P_Y)&= \frac{1}{2}\left[ \sum_{i=1}^r\ell_i^2 -1\right]- \frac{1}{2}\left( r_G + \frac{4}{3}d_Gh_G\right)+ \sum_{k=1}^N(2k-1)p_k \\
	}\label{eq:nhy}}
For example, the maximal puncture contributes
	\ba{
	n_v(P_{\text{max}}) =-\frac{1}{2}(N^2-1),\qquad n_h(P_{\text{max}})=0, \label{eq:max} 
	}
and the minimal puncture contributes
	\ba{
	n_v(P_{\text{min}}) =-\frac{1}{6}(4N^3-6N^2-N+3),\quad n_h(P_{\text{min}})=-\frac{1}{6} (4N^3-6N^2-4N). \label{eq:minans}
	}

An $SU(n_i)$ flavor group factor corresponds in the Young diagram to a nonzero difference of $n_i=\ell_i-\ell_{i+1}$ between the lengths of two rows. Then, the associated flavor central charge can be written
	\ba{
	k_{SU(\ell_i-\ell_{i+1})}=2\sum_{n\leq i}\ell_n.\label{eq:ky}
	}


\bibliographystyle{utphys}
\bibliography{bibliography}

\end{document}